\documentclass[twocolumn]{aastex631}

\usepackage{subfigure}
\usepackage{amsmath}
\usepackage{hyperref}
\usepackage{placeins}

\begin{document}

\title{Ly$\rm \alpha$ halos and UV continuum morphologies of Tadpole Galaxies at $z> 3$}

\author[0009-0008-2970-9845]{Manish Kataria}\thanks{E-mail: manish@iucaa.in}
\author[0000-0002-8768-9298]{Kanak Saha}\thanks{E-mail: kanak@iucaa.in}
\affiliation{Inter-University Centre for Astronomy and Astrophysics, Pune 411007, India.}
 
\begin{abstract}

Tadpole and clump-chain galaxies are a morphologically distinct population among high-redshift star-forming galaxies whose disturbed structures may influence the escape and propagation of Ly$\alpha$ photons. We investigate the Ly$\alpha$ and UV continuum properties of 12 tadpole galaxies in the redshift range of z $\sim$ 3 -- 5.5 identified in the Hubble Ultra Deep Field (HUDF) using deep MUSE observations. Accounting for their elongated morphologies, we construct surface brightness profiles and characterize the spatial extent of their Ly$\alpha$ emission. Extended Ly$\alpha$ halos are detected in 10 of the 12 galaxies, demonstrating that diffuse Ly$\alpha$ emission is common among tadpole systems. Approximately 40\% of the sample exhibits double-peaked Ly$\alpha$ profiles. While the effective radii ($\rm R_{e}$) of the Ly$\rm \alpha$ emission generally follow the spatial extent of the UV continuum, the Ly$ \rm \alpha$ halos are typically more symmetric and often exhibit spatial offsets from the stellar component. Some galaxies also display asymmetric and outflow-like Ly$\rm \alpha$ structures suggestive of anisotropic escape and complex radiative transfer effects. Together, these results suggest that the disturbed morphologies of tadpole galaxies may influence the transport of Ly$\rm \alpha$ photons and contribute to the formation of extended Ly$\rm \alpha$ halos in the circumgalactic medium.
\end{abstract}


\section{Introduction} \label{sec:intro}

Understanding how galaxies assemble and interact with their gaseous environment is a central goal of observational astronomy, particularly during the early epoch at redshift $z\ge 3$ when star formation and gas accretion are at their peak \citep{Madau_Dickinson2014, Tacconi_etal2020, Schreiber_etal2020}. High-resolution imaging from the Hubble Space Telescope (HST) has established that star-forming galaxies at these epochs are intrinsically compact, with rest-frame ultraviolet (UV) emission tracing young stellar populations on sub-kpc scales \citep{Ferguson_etal2004, Oesch_etal2010, Mosleh_etal2012, vanderWel_etal2014, Shibuya_etal2019}. This has now been extended beyond $z>6$ by the James Webb Space Telescope (JWST), allowing the measurement of galaxy UV sizes when they were extremely young \citep{Ferreira_etal2022, Chen_etal2023, Morishita_etal2024, Ormerod_etal2024}. On the other hand, integral field spectroscopic studies e.g., by the Multi Unit Spectroscopic Explorer (MUSE) on ESO/VLT, have revealed that Ly$\rm \alpha$ emission from galaxies at $z>3$ is often distributed over much larger spatial scales, forming diffuse halos that extend to several tens of kpc \citep{Steidel_etal2011, Zheng_etal2011, Momose_etal2014, Wisotzki_etal2016, Leclercq_etal2017}. The significant contrast between the compact UV emission, tracing the stellar distribution, and the extended Ly$\rm \alpha$ halos offers a unique probe of the physical connection between galaxies and their surrounding gas. In addition, tadpole galaxies offer a unique opportunity to investigate the relationship between Ly$\rm \alpha$ and UV spatial extents in morphologically disturbed systems.

The physical origin of extended Ly$\rm \alpha$ emission remains an open question \citep{Hayes2015}. As resonant transitions of neutral hydrogen gas, Ly$\rm \alpha$ photons undergo multiple scatterings within the interstellar and circumgalactic medium (ISM and CGM), allowing them to diffuse far from their production sites \citep{Adams1972, Harrington1973, Neufeld1990, Ahn_etal2001, Verhamme_etal2006, Dijkstra2014}. Radiative transfer effects alone can therefore generate extended emission even when the intrinsic source is compact. However, observational results indicate that Ly$\rm \alpha$ halos are nearly ubiquitous and can contribute a substantial fraction of the total Ly$\rm \alpha$ flux \citep{Leclercq_etal2017}, suggesting that additional processes may be in action, including the in-situ emission from cooling gas, fluorescence induced by ionizing radiation, and contributions from galactic outflows \citep{Cantalupo_etal2005, Dijkstra_etal2006, Dijkstra_Loeb_etal2009}.

A direct comparison of rest-frame UV and Ly$\rm \alpha$ sizes can provide a powerful diagnostic of this interplay. While UV emission traces sites of young star-formation, the Ly$\rm \alpha$ is sensitive to the distribution, kinematics, and ionization state of neutral hydrogen gas in and around the host galaxies. The ratio of the Ly$\rm \alpha$ to UV size therefore encodes crucial information about the radiative transfer, kinematics, and structure of the ISM and CGM in high-z galaxies. While a majority of high-z galaxies are compact in rest-frame UV, where the ratio $Ly{\rm \alpha}/UV > 1$, a minority of galaxies exhibit extended or highly asymmetric/elongated morphologies \citep{Abraham_etal1996}, especially the so-called tadpole galaxies characterized by bright star-forming head and diffuse tail \citep{ElmegreenD_etal2005, Straughn_etal2006}. These galaxies are likely in dynamically complex phases of evolution driven by mergers and clump migration, with actively ongoing mass accretion \citep{Kataria_etal2025}. In such systems, the basic assumption of a centrally concentrated Ly$\rm \alpha$ source may no longer hold, and the spatial distribution of Ly$\rm \alpha$ emission may instead be influenced by the underlying stellar geometry, kinematics, or anisotropic escape pathways \citep{Hayes_etal2013, Behrens_Braun2014, Leclercq_etal2017}.     

Investigating the relationship between the UV and Ly$\rm \alpha$ sizes and morphologies in these tadpole galaxies may provide a critical test for the current interpretation of Ly$\rm \alpha$ halos. If the Ly$\rm \alpha$ emission remains significantly extended than their UV stellar continuum even in these galaxies with extended star-formation, this would support a scenario where CGM plays a dominant role in shaping the Ly$\rm \alpha$ morphology \citep{Wisotzki_etal2016, Leclercq_etal2017}. Conversely, a closer correspondence between the UV and Ly$\rm \alpha$ emission would suggest that radiative transfer effects are more tightly coupled to the spatial distribution of young star-forming regions. Such comparisons are particularly insightful when performed galaxy-by-galaxy, thereby enabling exploration of correlations with morphology. In other words, the distinctive morphologies of tadpole galaxies provide a unique testbed for disentangling the physical processes underlying extended Ly$\rm \alpha$ emission \citep{Steidel_etal2011, Dijkstra2014}.
 
In this work, we present a spatially resolved study of UV and Ly$\rm \alpha$ for a sample of 12 tadpole galaxies at redshift $z>3$. By measuring their sizes and morphologies, we quantify the spatial decoupling between Ly$\rm \alpha$ emission and the stellar continuum, and investigate how this relationship depends on the internal structure of these galaxies. This approach provides a new perspective on the processes governing Ly$\rm \alpha$ escape, the role of the CGM, and the potential connection between host galaxy properties and their surrounding gas distribution.

Throughout this paper, we adopt a flat $\Lambda$ Cold Dark Matter $\rm (\Lambda CDM)$ cosmology \citep{White&Rees1978, Blumenthal_etal1984}, assuming $\rm H_{0} = 70~km~s^{-1}~Mpc^{-1},~\Omega_{M} = 0.3,~and~\Omega_{\Lambda} = 0.7$ for all scale calculations. Magnitudes, if present, are in the AB system \citep{Oke_Gunn1983}, and are used in the calculations presented in sec.~\ref{sec:flux_and_ew_lya}.

\section{data and source selection} \label{sec:data_source}

\subsection{Observational Data} \label{subsec:data}

In this study, we use observations obtained with the Multi Unit Spectroscopic Explorer (MUSE) in the Hubble Ultra Deep Field (HUDF), combining data from the HUDF MUSE Mosaic, the UDF-10 field, and the MUSE eXtremely Deep Field (MXDF) observations \citep{Bacon_etal2017, Bacon_etal2023}. These datasets provide spectroscopic coverage of the HUDF at different depths and spatial scales, enabling the study of faint, high-redshift galaxies.

The Mosaic observations consist of 9 individual pointings that together cover approximately $\sim$90\% of the HUDF area. Each pointing has an integration time of $\sim$10~hrs and a total field of view of 9.92 arcmin$^2$ from all observations, providing relatively wide spectroscopic coverage across the field. The UDF-10 field is located within the Mosaic footprint and covers a smaller region of $\sim$1.15 arcmin$^2$, but with a significantly deeper exposure time of $\sim$31~hrs.

The deepest observations are provided by the MUSE eXtremely Deep Field (MXDF), which reaches a total integration time of $\sim$140~hrs. The MXDF is co-spatial with the HUDF's extremely deep region \citep{Illingworth_etal2013}, enabling extremely sensitive spectroscopic measurements of faint galaxies and extended emission. Together, these datasets provide a combination of wide-area coverage and ultra-deep spectroscopy, making them particularly well-suited for studying faint line emission from distant galaxies with weak stellar continuum. 

The MUSE instrument operates in an optical wavelength range of $\sim$ 4750 -- 9350~\text{\AA} with a linear sampling of 1.25~\text{\AA}, hence the $\rm Ly\alpha$ emitters in the redshift range of z $\sim$ 3 -- 6.7 can be observed with the help of MUSE. Along with the increased sensitivity and large field of view ($\rm 1^{\prime}\times1^{\prime}$) from the previous instruments, we can now observe individual galaxy $\rm Ly\alpha$ emission \citep{Bacon_etal2015, Drake_etal2017}.   

In this work, we use publicly available high-resolution imaging from the Hubble Space Telescope (HST), obtained from the Hubble Legacy Archive (HLA) and the JWST Advanced Deep Extragalactic Survey (JADES; PI: Daniel Eisenstein and Nora Luetzgendorf) \citep{Eisenstein_etal2026}, respectively. These datasets provide high–spatial resolution imaging that complements the spectroscopic observations used in this study. To trace the rest-frame UV continuum emission of the galaxies, we primarily use HST/WFPC2 F814W imaging. This filter is chosen because it lies redward of the $\rm Ly\alpha$ forest for the redshift range of our sources, thereby minimizing attenuation from intergalactic neutral hydrogen and providing a cleaner measurement of the UV continuum emission.

\subsection{Sample selection} \label{subsec:source}

\begin{figure*}
    \centering
    \includegraphics[width = 6in]{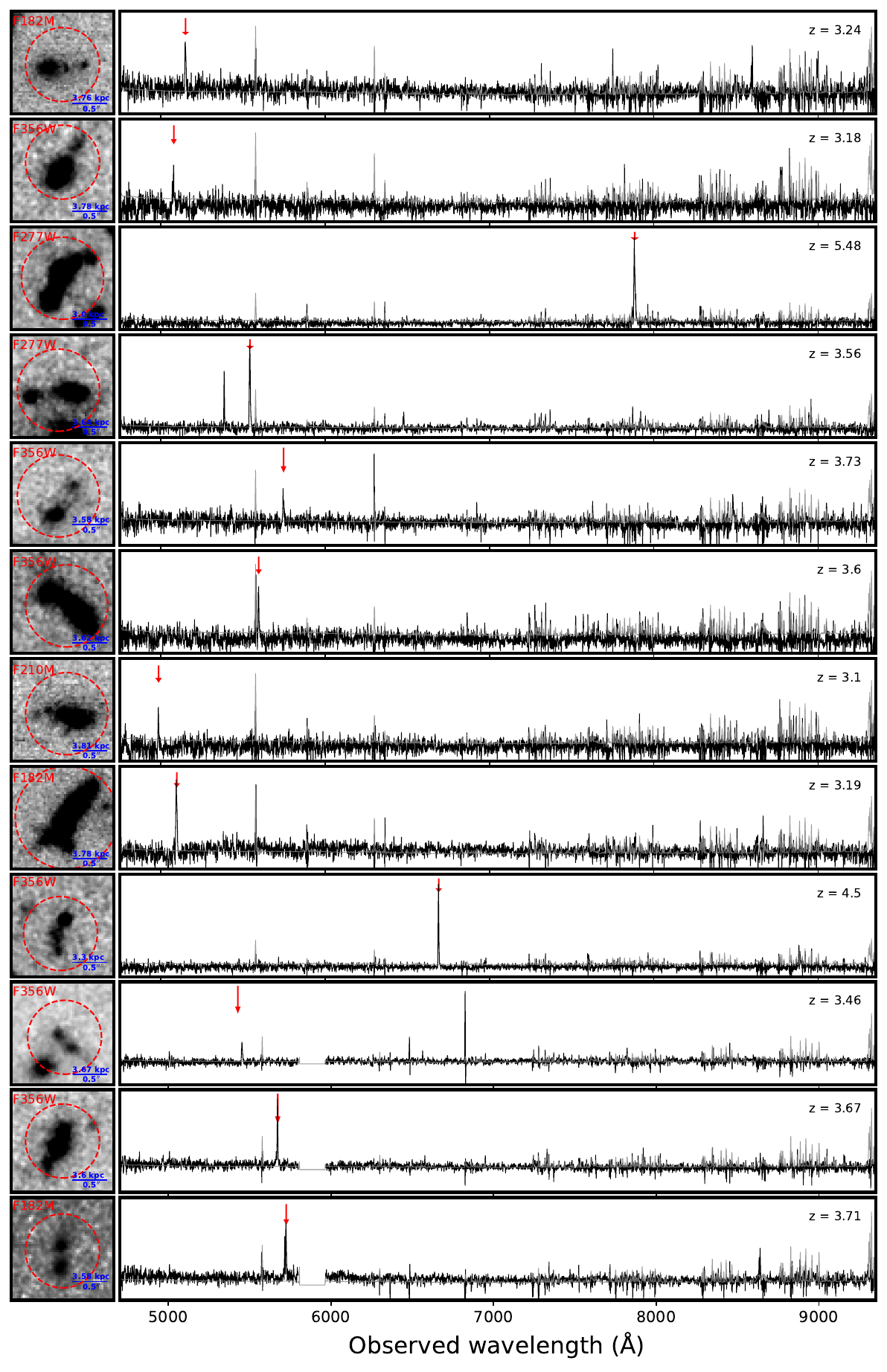}
    \caption{Sample of high-z $\rm Ly\alpha$ emitting Tadpole/chain galaxies. The left panel shows the JWST/NIRCAM band image (filter is mentioned in the top left corner of each image with red color) of the galaxy within a dashed red circle, with scale in kpc (at the redshift of the galaxy) and arcsec shown in blue. The right panels show the corresponding MUSE spectra in black and the error in gray, with the $\rm Ly\alpha$ emission line marked with the downward pointing arrow at the observed wavelength.}
    \label{fig:tpg_lya_spec}
\end{figure*}

\begin{figure*}[ht!]
\centering
\subfigure{\includegraphics[width = 2.3in]{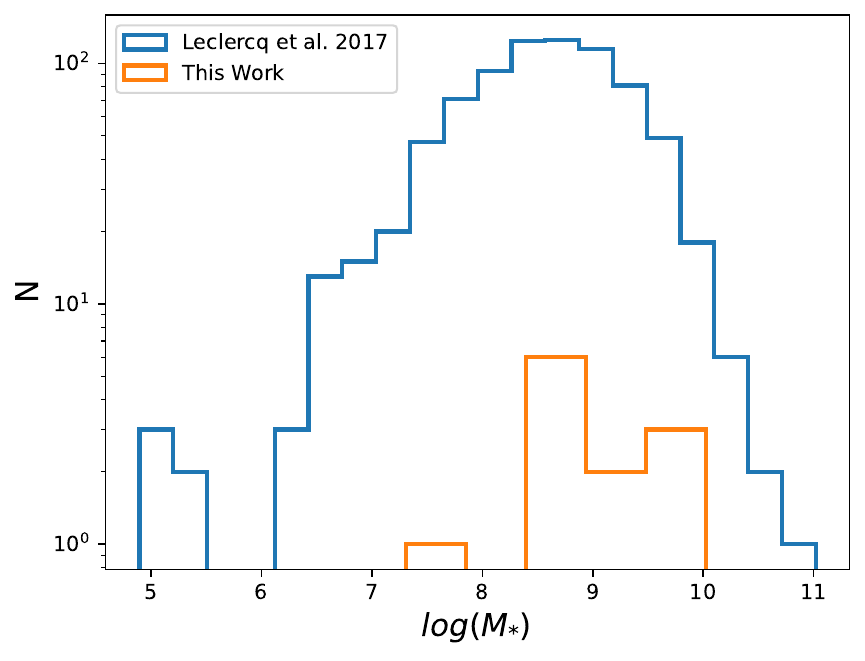}}
\hfill
\subfigure{\includegraphics[width = 2.3in]{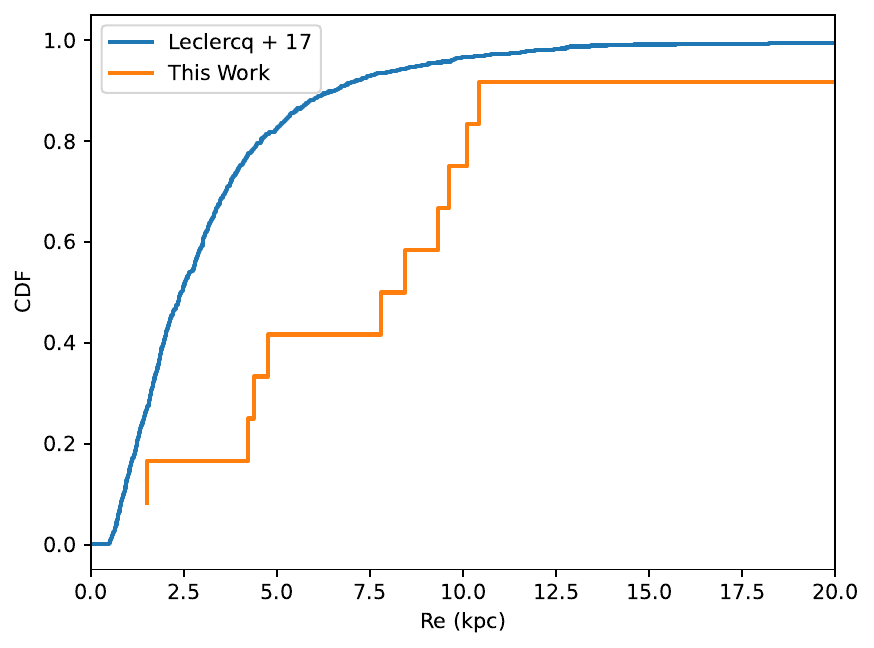}}
\hfill
\subfigure{\includegraphics[width = 2.3in]{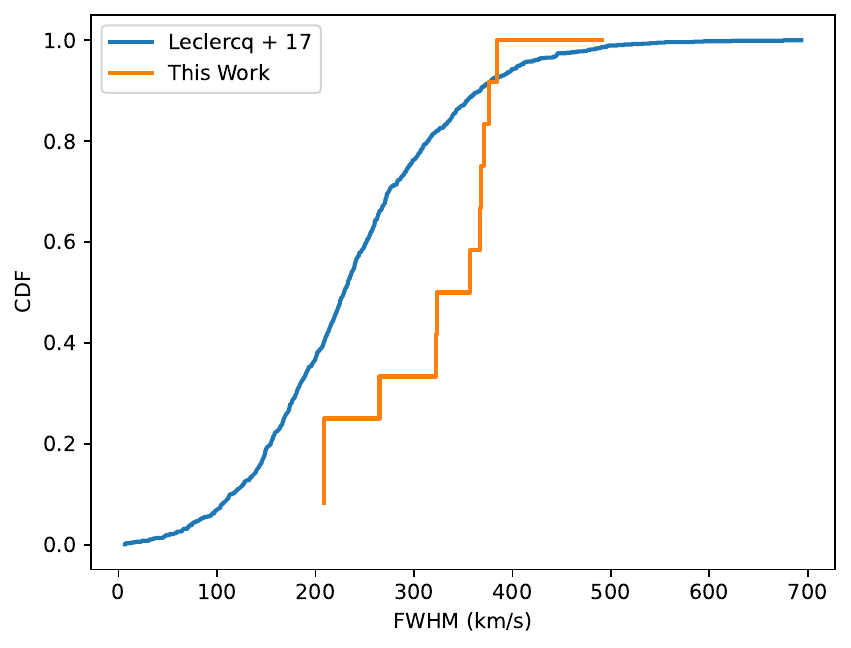}}
\caption{Comparison of the galaxy sample with \cite{Leclercq_etal2017}. The left panel shows the distribution of stellar mass for the GOODS-South galaxies in our sample. The middle and right panels show the cumulative distributions for all galaxies in the $\rm Ly\alpha$ halo scale length and the red peak of the emission.}
\label{fig:sample_comparison}
\end{figure*}

Tadpole and chain-like elongated galaxies provide an excellent opportunity to study the spatially resolved properties of galaxies at high redshift, mainly due to their extended, elongated, and clumpy morphologies. These systems are often interpreted as either merging galaxies or edge-on disks still in the process of settling \citep{Sanchez_Almeida_etal2013}. Their clumpy structures are commonly attributed to strong dynamical instabilities within gas-rich disks. The first systematic identification of such tadpole and clump-chain galaxies was made in the Hubble Ultra Deep Field by \cite{Cowie_etal1995, ElmegreenD_etal2005}. 

In the local Universe, the Kiso survey \citep{Miyauchi-Isobe_etal2010} found that tadpole galaxies constitute only about 0.2\% of the star-forming galaxy population \citep{Elmegreen_DM_etal2010}, in contrast to the much higher fraction of $\sim10\%$ observed in the HUDF \citep{Elmegreen_DM_etal2007, Elmegreen&Elmegreen2010}. This difference has been interpreted as evidence that tadpole galaxies represent a transient phase in the formation and evolution of disk galaxies. Furthermore, local tadpole galaxies show a strong overlap with the rare population of extremely metal-poor (XMP \citep{Kunth&Ostlin2000}) galaxies, while many XMP galaxies themselves exhibit tadpole-like morphologies \citep{Papaderos_etal2008, Morales-Luis_etal2011, Filho_etal2013}. This association suggests that tadpole galaxies may be among the youngest systems in the local Universe. Additional support for this picture comes from the observed metallicity drops at the locations of their bright star-forming heads, which have been interpreted as signatures of ongoing accretion of pristine gas through cosmic filaments \citep{Sanchez_etal2014, Sanchez_etal2015}.

We construct a sample of tadpole galaxies from the catalog of \citet{Straughn_etal2006}. The catalog contains 165 galaxies identified as tadpoles, of which 23 have photometric redshifts $\rm z > 3$. We cross-matched the full catalog with the MUSE emission-line catalog of \cite{Bacon_etal2017}, yielding 24 galaxies with spectroscopic redshifts $\rm z > 3$ within a matching radius of $1^{\prime\prime}$. Among these, 7 galaxies overlap with the subset of 23 objects having photometric redshifts above 3. The remaining matches arise from discrepancies between the photometric and spectroscopic redshift estimates. Some galaxies with photometric redshifts $\rm z < 3$ were found by MUSE to lie at $\rm z > 3$, while some objects with photometric redshifts $\rm z > 3$ were spectroscopically confirmed to be at lower redshifts. We subsequently visually inspected the MUSE spectra and excluded sources with low signal-to-noise ratios or significant contamination from nearby objects. This procedure yielded a final sample of 12 galaxies, listed in Table~\ref{tab:muse_lya_sample} and shown in Figure~\ref{fig:tpg_lya_spec}. Among these, 7 galaxies are located in the HUDF MUSE Mosaic field, 2 in the UDF-10 field, and 3 in the MUSE eXtremely Deep Field (MXDF). The sample consists of 6 galaxies with single-peaked $\rm Ly\alpha$ profiles, 5 with double-peaked profiles, and one galaxy showing tentative evidence for a triple-peaked profile.
 
The galaxies in our sample span a range of stellar masses - $\rm 7.3 <log\ M_{*}(M_{\odot}) < 10$ taken from \cite{Santini_etal2015} as shown in Fig.~\ref{fig:sample_comparison} in comparison with GOODS-South galaxies in \citep{Leclercq_etal2017}.

{\renewcommand{\arraystretch}{1.5}
\begin{table*}[ht]
    \centering
    \begin{tabular}{|l|c|c|c|c|c|c|c|c|}
    \hline
        \#ID & RA\ (Deg.) & Dec\ (Deg.) &  z & $\rm R_{eff, UVC}\ (kpc)$ & $\rm R_{eff, Ly\alpha, 1c}\ (kpc)$ & $\rm R_{eff, halo}\ (kpc)$ & $\rm F_{Ly\alpha}$ & $\rm log(L_{Ly\alpha})$\\
    \hline\hline
        $1^{s}$ & 53.12548823 & -27.78822322 &  3.237545 & $4.14_{-0.57}^{+0.57}$ & $7.54_{-0.99}^{+1.13}$ & $>9.33$ & $8.9\pm 0.2$ & 41.92\\
        $2^{d}$ & 53.13273408 & -27.79503827 & 3.178914 & $8.85_{-3.03}^{+2.74}$ & $8.76_{-1.13}^{+1.35}$ & $>8.44$ & $9.1\pm0.2$ & 41.91\\
        $3^{d}$ & 53.13857927 & -27.7902169 & 5.484836 & $1.89_{-0.38}^{+0.43}$ & $3.03_{-0.12}^{+0.12}$ & $7.79_{-1.74}^{+2.83}$ & $60.4\pm1.9$ & 43.29\\
        $4^{d}$ & 53.15756862 & -27.76470463 & 3.56046 & $2.65_{-0.63}^{+0.61}$ & $3.19_{-0.16}^{+0.17}$ & $4.76_{-1.31}^{+7.58}$ & $35.1\pm0.4$ & 42.61\\
        \textbf{$5^{s}$*} & 53.16055594 & -27.77109065 & 3.728073 & $2.05_{-1.50}^{+3.36}$ & $3.03_{-0.49}^{+0.54}$ & $4.22_{-1.09}^{+4.39}$ & $4.8\pm0.2$ & 41.80\\
        $6^{s}$ & 53.16086508 & -27.80112209 & 3.603717 & $3.85_{-0.71}^{+0.71}$ & $10.33_{-1.12}^{+1.30}$ & $10.43_{-1.26}^{+1.85}$ & $13.2\pm0.4$ & 42.20\\
        $7^{s}$ & 53.16222974 & -27.81576889 & 3.102812 & $2.67_{-0.30}^{+0.30}$ & $22.78_{-6.76}^{+14.38}$ & $>21.55$ & $6.9\pm0.4$ & 41.76\\
        $8^{d}$ & 53.16492565 & -27.76512153 & 3.194288 & $2.93_{-0.14}^{+0.12}$ & $13.54_{-1.27}^{+1.47}$ & $22.47_{-5.91}^{+13.31}$ & $28.9\pm0.8$ & 42.41 \\
        \textbf{$9^{s}$*} & 53.16573139 & -27.77171562 & 4.504603 & $1.56_{-1.12}^{+2.96}$ & $1.52_{-0.08}^{+0.09}$ & $10.10_{-2.51}^{+3.42}$ & $17.6\pm0.5$ & 42.56\\
        \textbf{$10^{s}$**} & 53.16629394 & -27.78223589 & 3.464829 & $1.19_{-0.85}^{+1.95}$ & $4.00_{-0.59}^{+0.69}$ & $4.37_{-0.76}^{+1.29}$ & $2.1\pm0.1$ & 41.37\\
        \textbf{$11^{d}$**} & 53.16929757 & -27.77810572 & 3.666497 & $1.21_{-0.85}^{+1.38}$ & $1.09_{-0.17}^{+0.20}$ & $1.51_{-0.76}^{+1.29}$ & $5.9\pm0.1$ & 41.87\\
        \textbf{$12^{t}$**} & 53.17418858 & -27.78992382 & 3.709646 & $13.13_{-3.68}^{+4.28}$ & $2.77_{-0.09}^{+0.10}$ & $>2.84$ & $11.8\pm0.2$ & 42.18\\
        
    \hline    
    \end{tabular}
    \caption{Positions (RA \& Dec.) of $\rm Ly\alpha$ detected sources and their redshifts (z). The s, d, and t superscripts on the \#IDs are there to represent the line morphology of single, double and triple peak respectively. The $\rm R_{eff, UVC}$ is the effective radius from the 2D Sersic fit to the PSF convolved image of the UV continuum in the HST band, the $\rm R_{eff, Ly\alpha, 1c}$ is the effective radius from the single component Sersic fit to the $\rm Ly\alpha$ narrowband image. $\rm F_{Ly\alpha}$ is the total $\rm Ly\alpha$ flux in $\rm 10^{-18} erg s^{-1} cm^{-2}$ measured using the curve of growth method from narrow band images. The $\rm log(L_{Ly\alpha})$ is the $\rm log_{10}$ of the $\rm Ly\alpha$ Luminosity. }

    \vspace{0.2cm}
    \raggedright
    $^{\textbf{*}}$ The data was taken from UDF-10 MUSE observation. \\
    $^{\textbf{**}}$ The data was taken from MXDF MUSE observation. \\
    Note: If not marked, then the data was from MUDF 
    \label{tab:muse_lya_sample}
\end{table*}
}

\section{\texorpdfstring{Ly$\alpha$\ }\ narrow band images} \label{sec:lya_nb_img}

With IFU data such as MUSE, emission-line sources can be efficiently detected even when their continuum emission is extremely faint. For the Ly$\alpha$ emitters in our sample, we extract cube cutouts of $8.2^{\prime\prime} \times 8.2^{\prime\prime}$ from the MUSE datacube. To isolate the Ly$\alpha$ emission, we select a spectral window of 150~$\text{\AA}$ centered on the Ly$\alpha$ line, following a procedure similar to \cite{Drake_etal2017}. By restricting each galaxy’s cube to a small spatial and spectral region, we significantly reduce contamination from nearby sources and overlapping spectral features prior to continuum subtraction.

We perform the continuum subtraction following the methodology described in \cite{Herenz&Wisotzki2017}. For each cube cutout, a median filter with a width of 150 pixels is applied along the spectral axis to estimate the underlying continuum emission while suppressing narrow spectral features associated with emission lines. The resulting median-filtered cube serves as a model of the continuum component. This continuum model is subsequently subtracted from the original data cube, producing a continuum-subtracted cube in which the emission-line signal is isolated. The resulting continuum-free data cubes are then used for the extraction and analysis of the Ly$\rm \alpha$ emission associated with each galaxy.

The continuum-subtracted emission-line cube is then collapsed over the wavelength interval indicated by the faint red band in the middle panel of \ref{fig:Lya_halo_spec_rgb}. Due to the small number of galaxies in our sample, this wavelength range is selected for each source by visual inspection of the extracted spectrum, ensuring that the full extent of the Ly$\rm \alpha$ emission is included while minimizing the contribution from noise-dominated spectral channels.

\section{Preparation of the UV continuum images} \label{sec:uvc_img}

The rest-frame ultraviolet (UV) emission in galaxies primarily traces light from young, massive stars and is therefore a robust indicator of recent star-formation activity \citep{Kennicutt1998, Calzetti2013}. We use UV continuum images for our galaxy sample from the Hubble Legacy Archive (HLA; \cite{Whitaker_etal2019}). For all sources except one, we adopt the WFC/F814W filter, while for source \#ID = 3 (z = 5.48), we use the WFC3\_IR/F105W filter to better probe the rest-frame UV emission at higher redshift. The F814W filter is minimally affected by Lyman forest attenuation at the redshifts of our sample, making it more suitable than bluer filters such as F606W, which lie closer to the rest-frame $\sim1500~\text{\AA}$ and are more susceptible to absorption \citep{Madau1995, Inoue_etal2014}. We generate image cutouts of size $8.2^{\prime\prime}\times8.2^{\prime\prime}$ centered on each target galaxy. Segmentation maps are then constructed from the UV continuum images to identify and isolate sources within the field. Owing to the high spatial resolution of the HST, nearby contaminants can be effectively separated from the target galaxy. Using the segmentation maps, we mask all neighboring sources within the cutout region. The masked regions are subsequently filled by sampling from a Gaussian background model derived from the surrounding sky in the cutout. This procedure is iteratively repeated to ensure that no residual contamination from nearby sources remains.

After cleaning the UV continuum (UVC) images, we process them to ensure a direct comparison with the $\rm Ly\alpha$ narrowband images. First, we rebin the HST images from their native pixel scale of $0.06^{\prime\prime}\mathrm{pix}^{-1}$ to match the MUSE pixel scale of $0.2^{\prime\prime}\mathrm{pix}^{-1}$. To account for the difference in spatial resolution, we construct wavelength-dependent point spread function (PSF) cubes using the \texttt{mpdaf} package \citep{Bacon_etal2016, Piqueras_etal2017}. The PSF modeling incorporates the wavelength-dependent full width at half maximum (FWHM$(\lambda)$) for different MUSE datasets, along with Moffat profile parameters, adopting $\beta_{\mathrm{MOSAIC,\ UDF\mbox{-}10}} = 2.8$ and $\beta_{\mathrm{MXDF}} = 2.123$ from \cite{Bacon_etal2017}. The PSF cubes are then restricted to the same wavelength range used to generate the $\rm Ly\alpha$ narrowband images. A corresponding 2D PSF is obtained by averaging the cube along the wavelength axis within this range. Finally, the cleaned and rebinned HST UVC images are convolved with these PSFs to match the MUSE spatial resolution, enabling a consistent comparison between the UV continuum and $\rm Ly\alpha$ emission.

\section{Surface brightness profiles} \label{sec:sb_profile}

\begin{figure*}[ht!]
\centering

\subfigure{\includegraphics[width = 1.67in, clip, trim = -0cm -2cm 0cm 0cm]{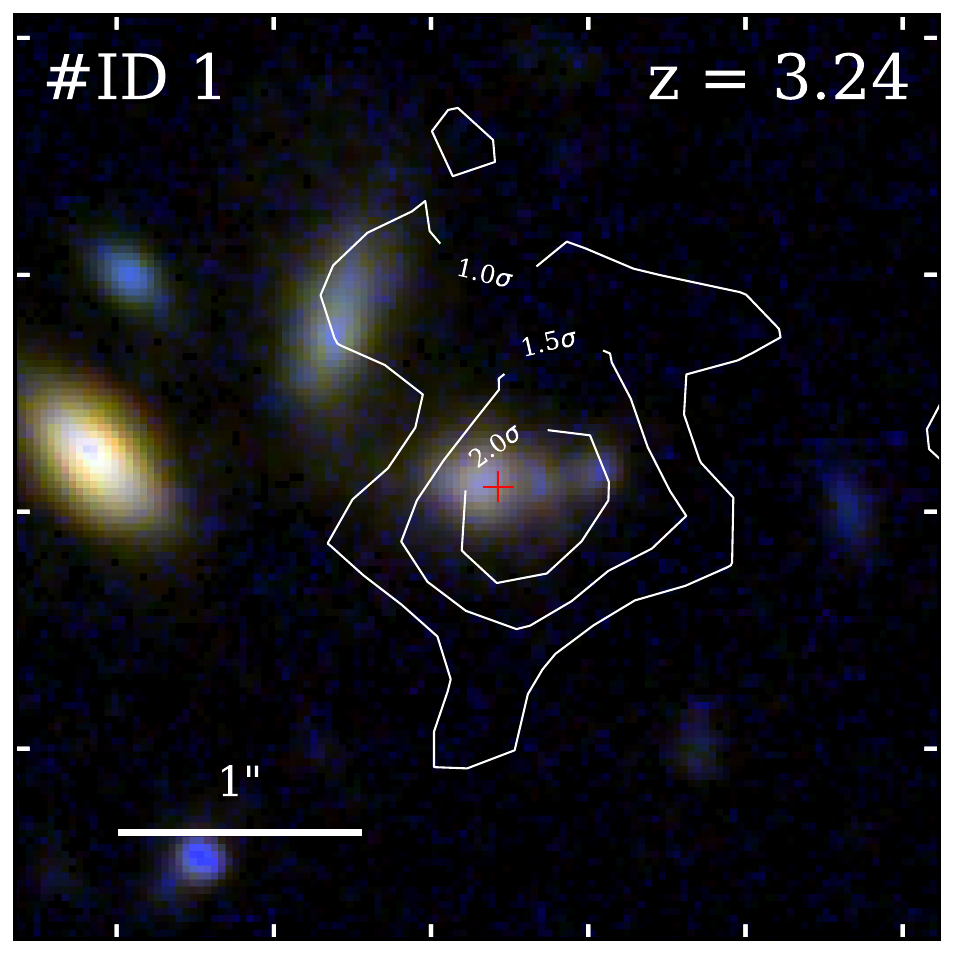}}
\hfill
\subfigure{\includegraphics[width = 2.3in, clip, trim = 0cm -0.5cm 0cm 0cm]{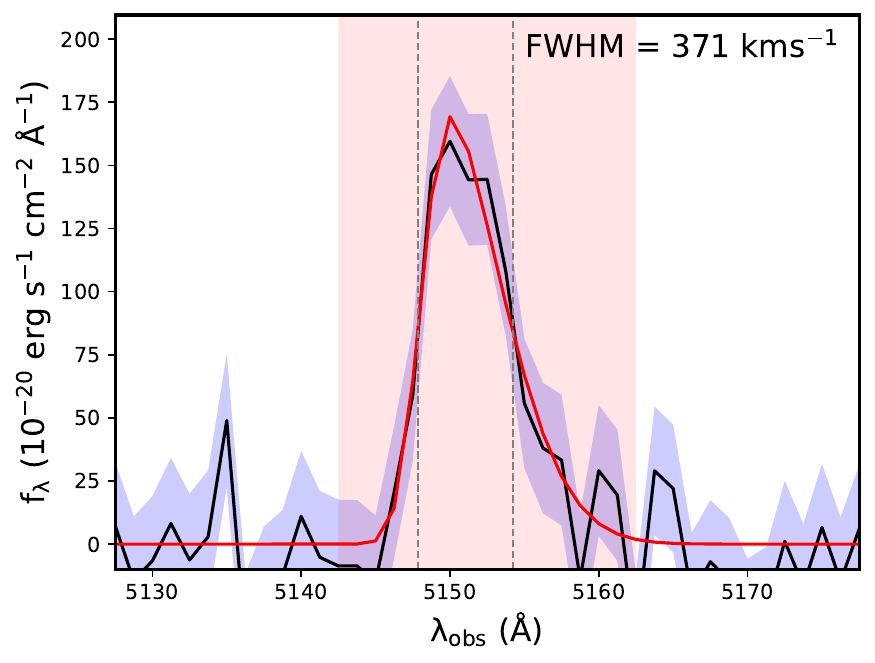}}
\hfill
\subfigure{\includegraphics[ width = 3in]{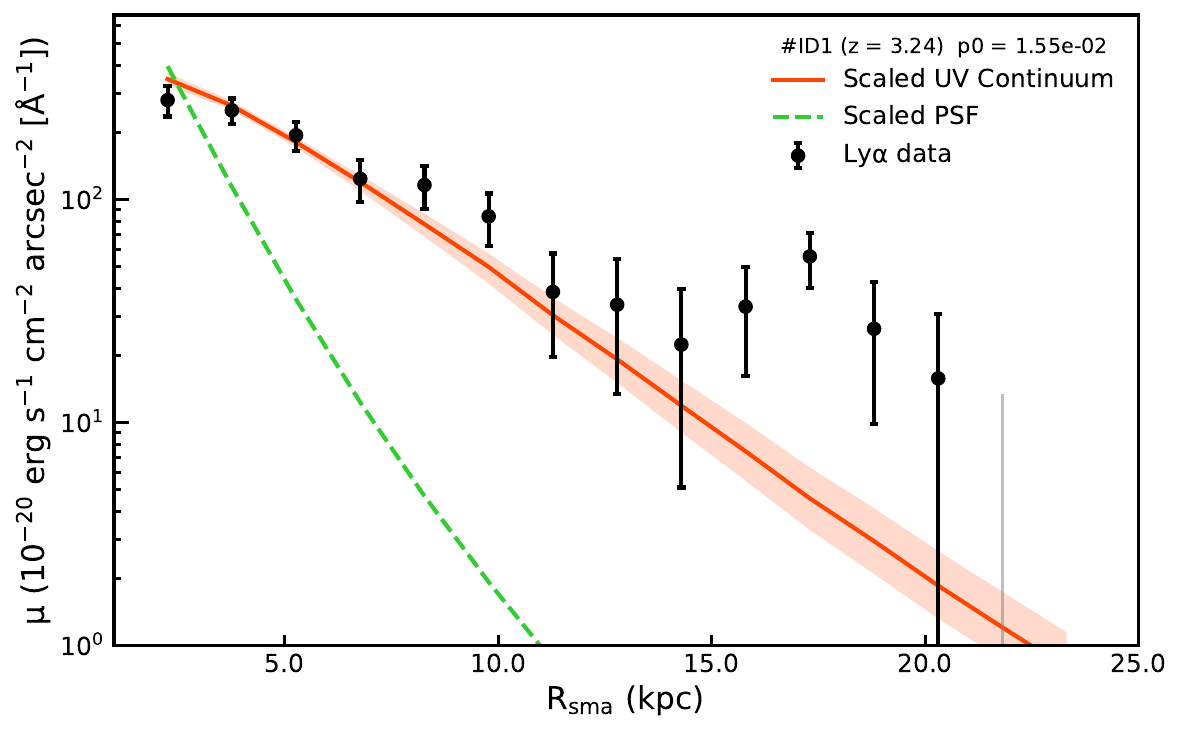}}\\

\vspace{-0.5cm}

\subfigure{\includegraphics[width = 1.67in, clip, trim = 0cm -2cm 0cm 0cm]{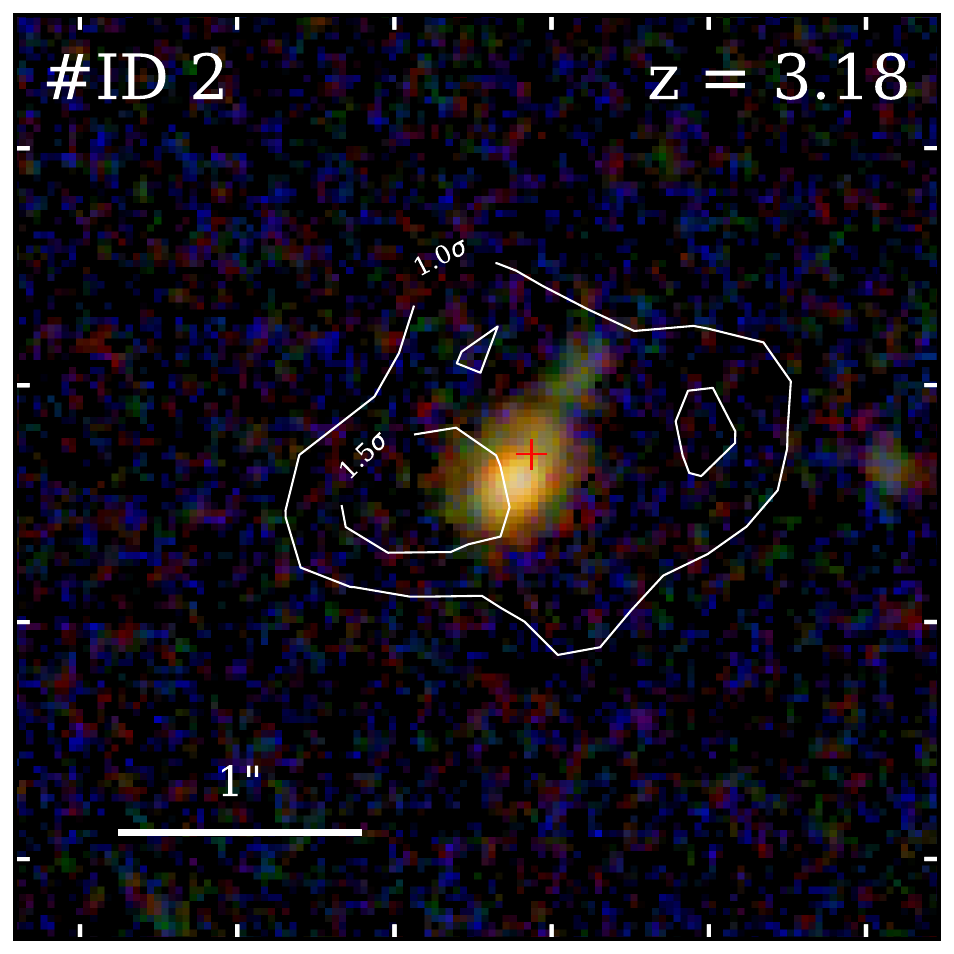}}
\hfill
\subfigure{\includegraphics[width = 2.3in, clip, trim = 0cm -0.5cm 0cm 0cm]{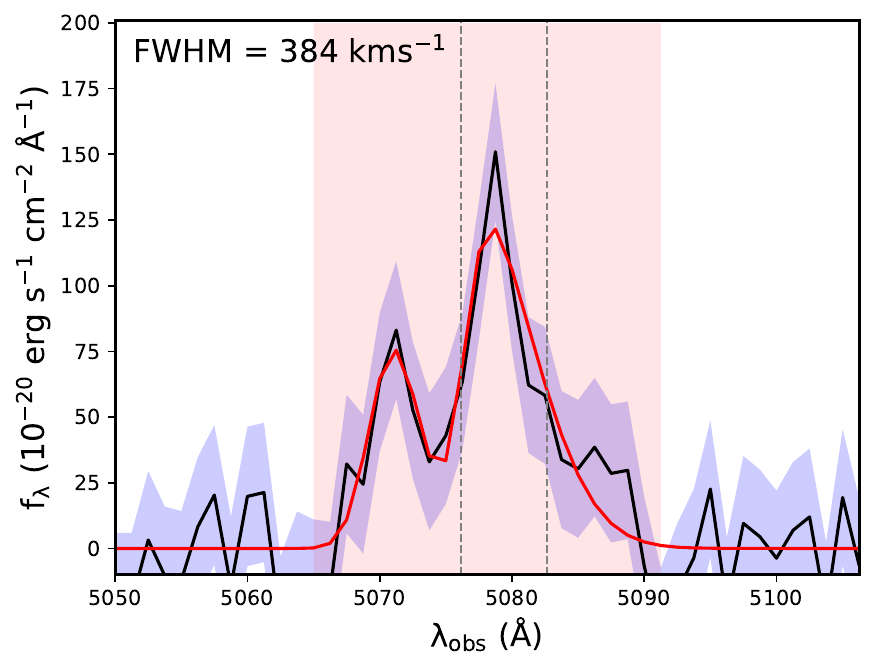}}
\hfill
\subfigure{\includegraphics[ width = 3in]{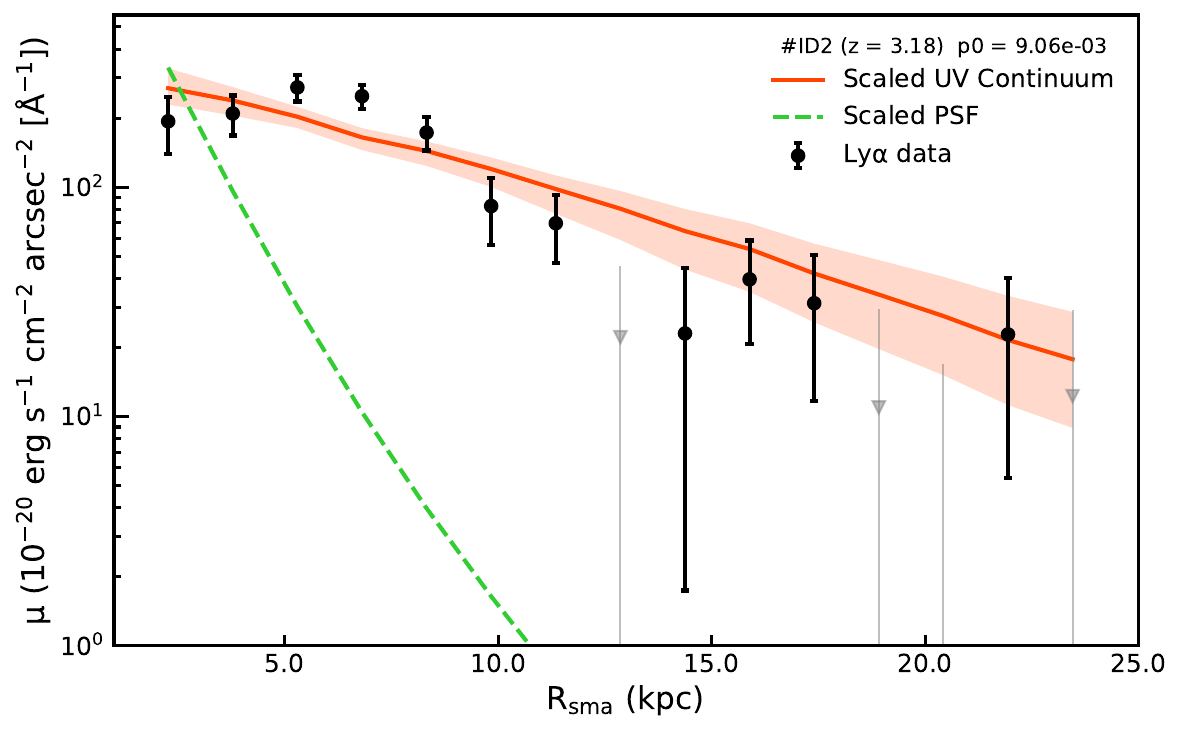}}\\

\vspace{-0.5cm}

\subfigure{\includegraphics[width = 1.67in, clip, trim = 0cm -2cm 0cm 0cm]{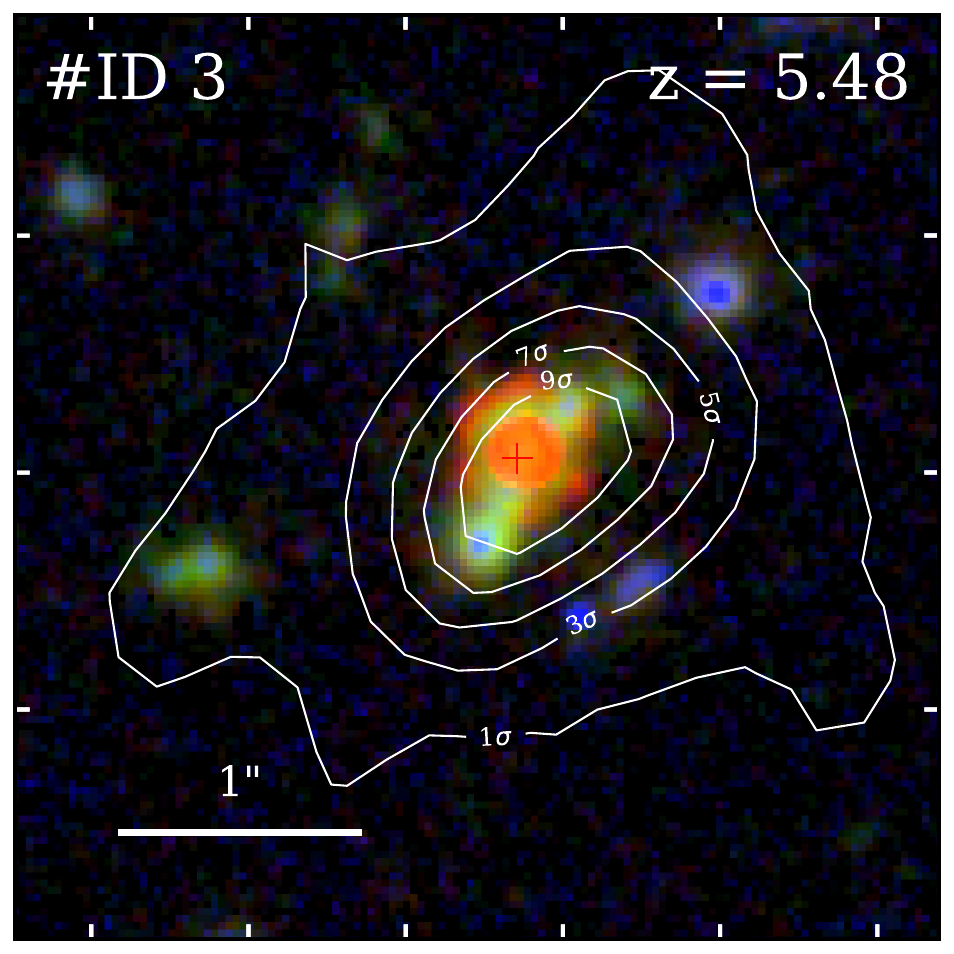}}
\hfill
\subfigure{\includegraphics[width = 2.3in, clip, trim = 0cm -0.5cm 0cm 0cm]{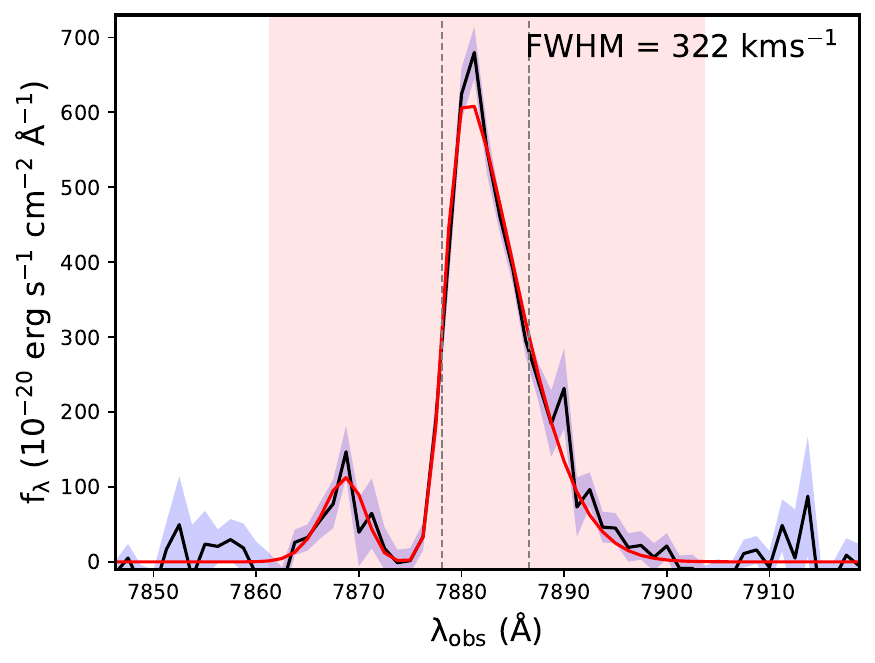}}
\hfill
\subfigure{\includegraphics[ width = 3in]{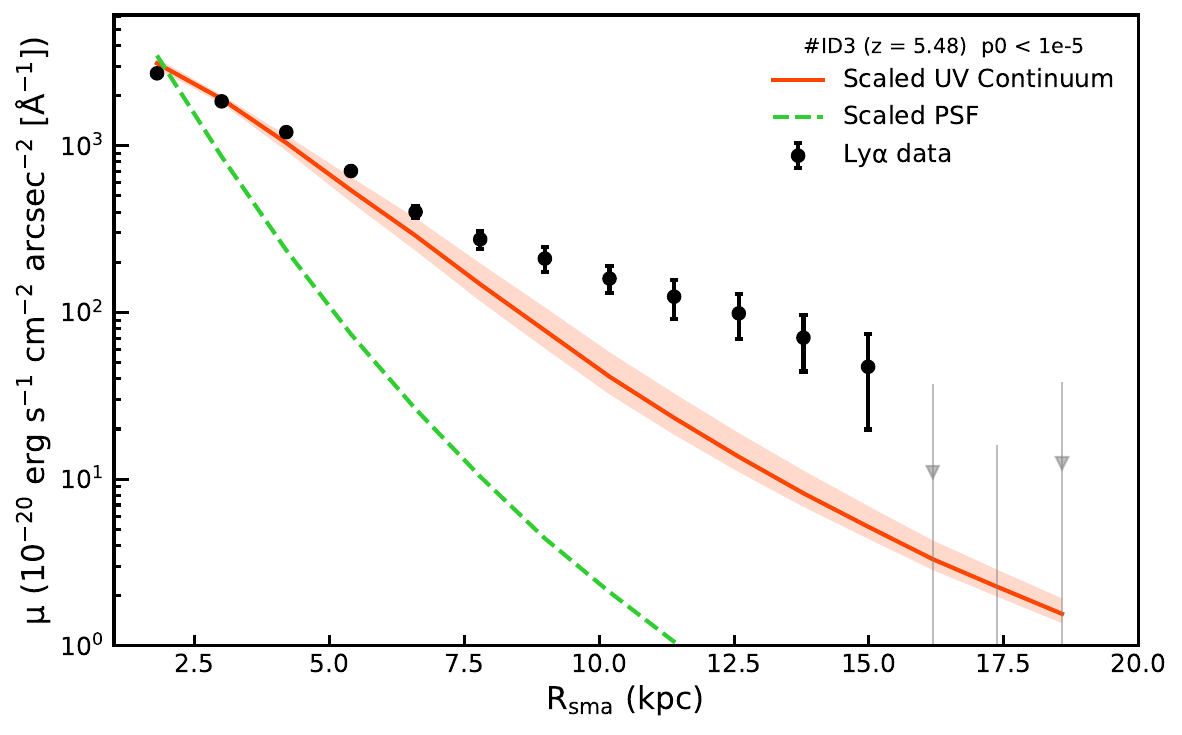}}\\

\caption{\textbf{Left:} The panel shows the JWST RGB image of the Tadpole/Chain galaxy analyzed in this work. The white contours show the $\rm Ly\alpha$ emission from MUSE, overlaid on the galaxy, with the contour levels specified inline in the contours themselves. At the top-left, the IDs of each galaxy are mentioned with the redshift mentioned in the top-right, and at the bottom-left of each image, we show the $1^{\prime\prime}$ scale as a white line segment. \textbf{Middle:} Panel shows the $\rm Ly\alpha$ emission line extracted from MUSE in black with the errors shown in faint blue color. The red line shows the asymmetric Gaussian fit to the observed emission line, the faint red vertical band represents the wavelength window considered for the construction of the $\rm Ly\alpha$ narrow band images, with the FWHM of the red peak mentioned at the top of each plot in $\rm kms^{-1}$. \textbf{Right:} This panel shows the Observed 1D $\rm Ly\alpha$ profiles measured from the narrow band images (see sec.~\ref{sec:lya_nb_img}) shown in black datapoints with the errorbars, the gray data points show the datapoints with SNR$<$1 or a negative flux. The orange line is the MCMC line fit to the 1D UV continuum profile, with the orange band showing the $\rm 16^{th}-84^{th}$ percentile confidence intervals. This UVC profile is scaled by a constant scale factor to best match the $\rm Ly\alpha$ profiles, since the flux units differ. The green dashed line represents the MUSE PSF for each narrow-band image within that wavelength. The p-value shown in each plot is the probability under the null hypothesis that the UV profile explains the observed $\rm Ly\alpha$ profile.}
\label{fig:Lya_halo_spec_rgb}
\end{figure*}

\begin{figure*}[ht!]
\centering
\addtocounter{figure}{-1}

\subfigure{\includegraphics[width = 1.67in, clip, trim = 0cm -2cm 0cm 0cm]{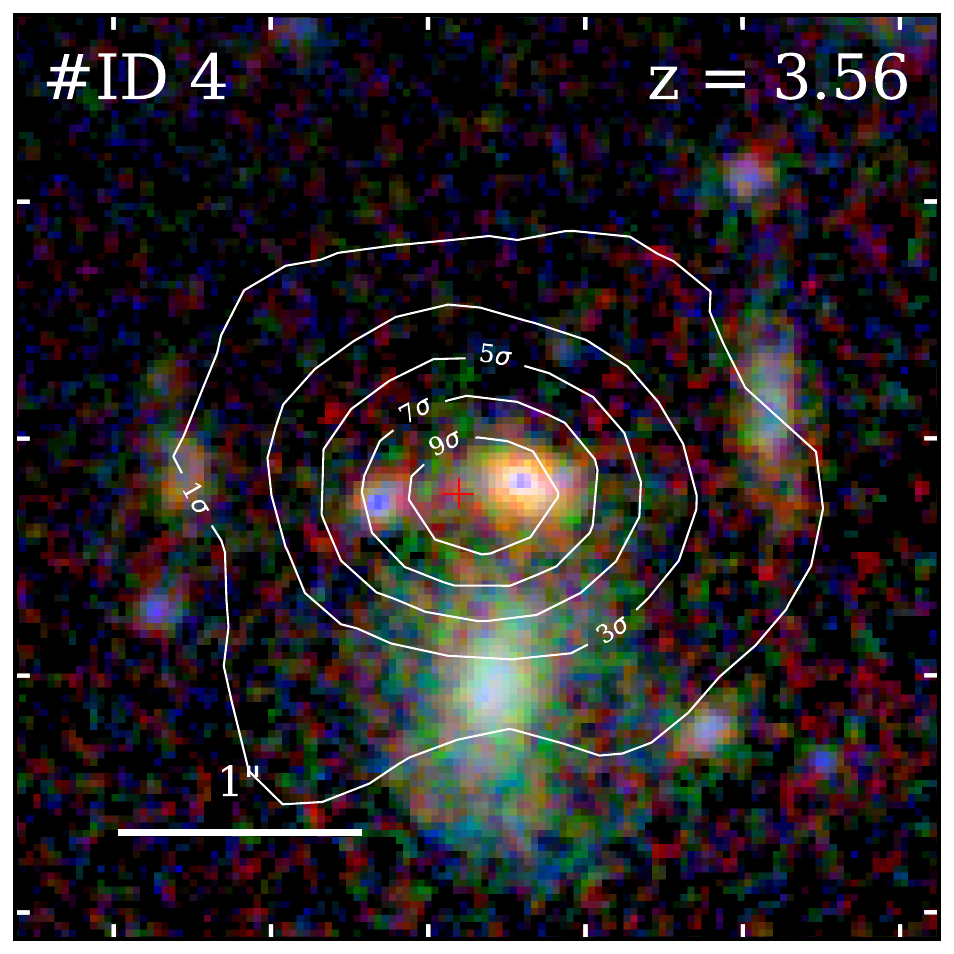}}
\hfill
\subfigure{\includegraphics[width = 2.3in, clip, trim = 0cm -0.5cm 0cm 0cm]{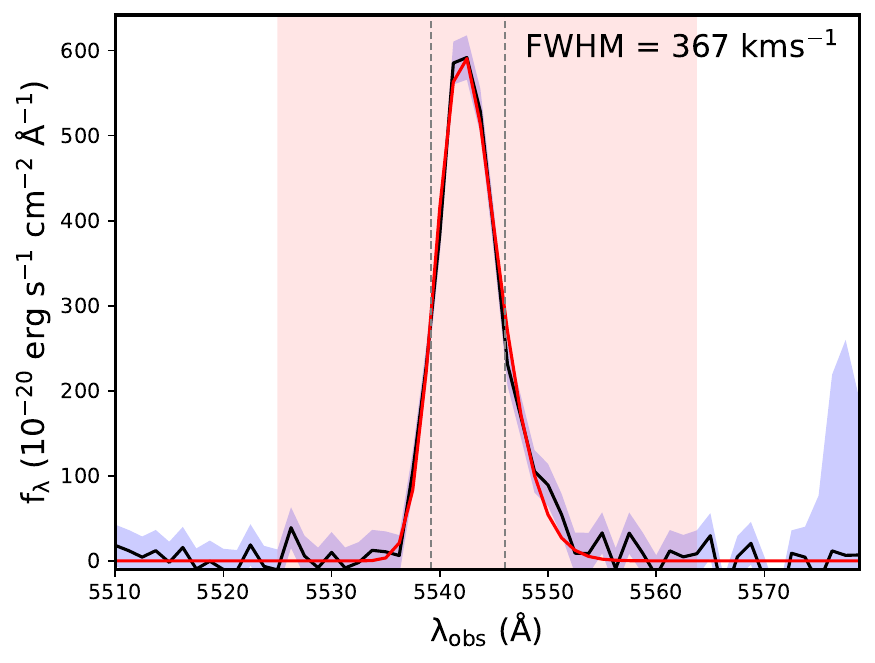}}
\hfill
\subfigure{\includegraphics[ width = 3in]{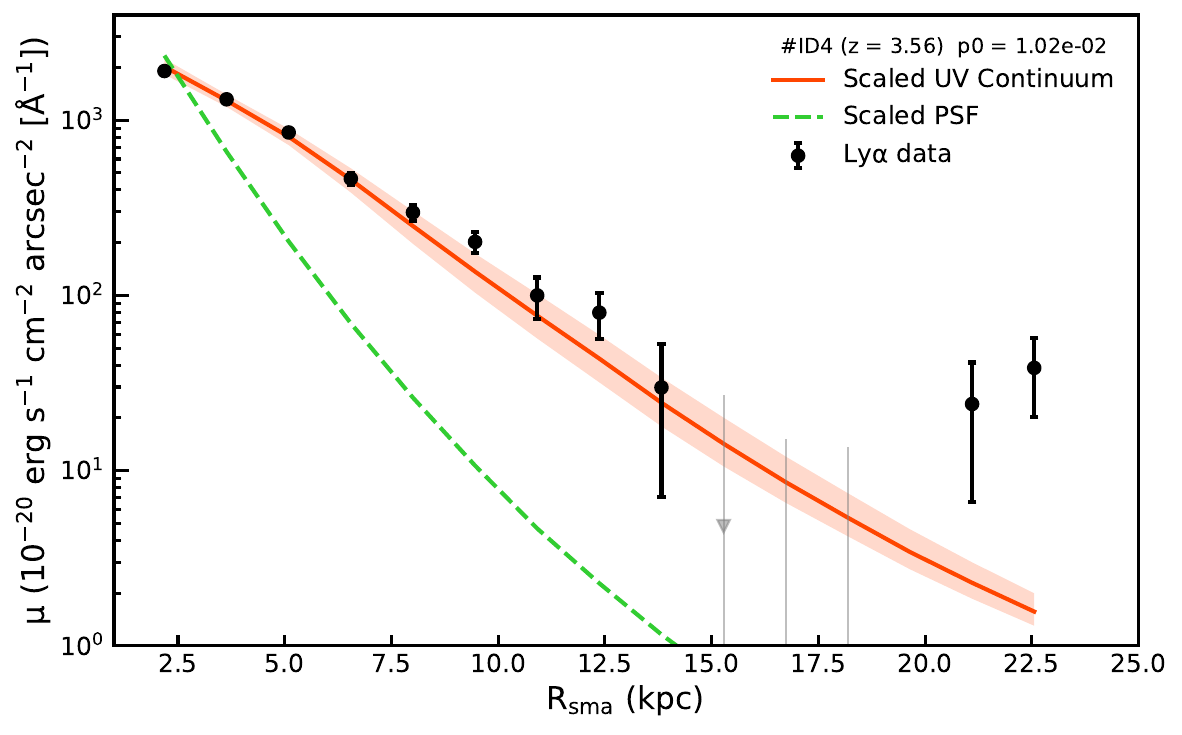}}\\

\vspace{-0.5cm}

\subfigure{\includegraphics[width = 1.67in, clip, trim = 0cm -2cm 0cm 0cm]{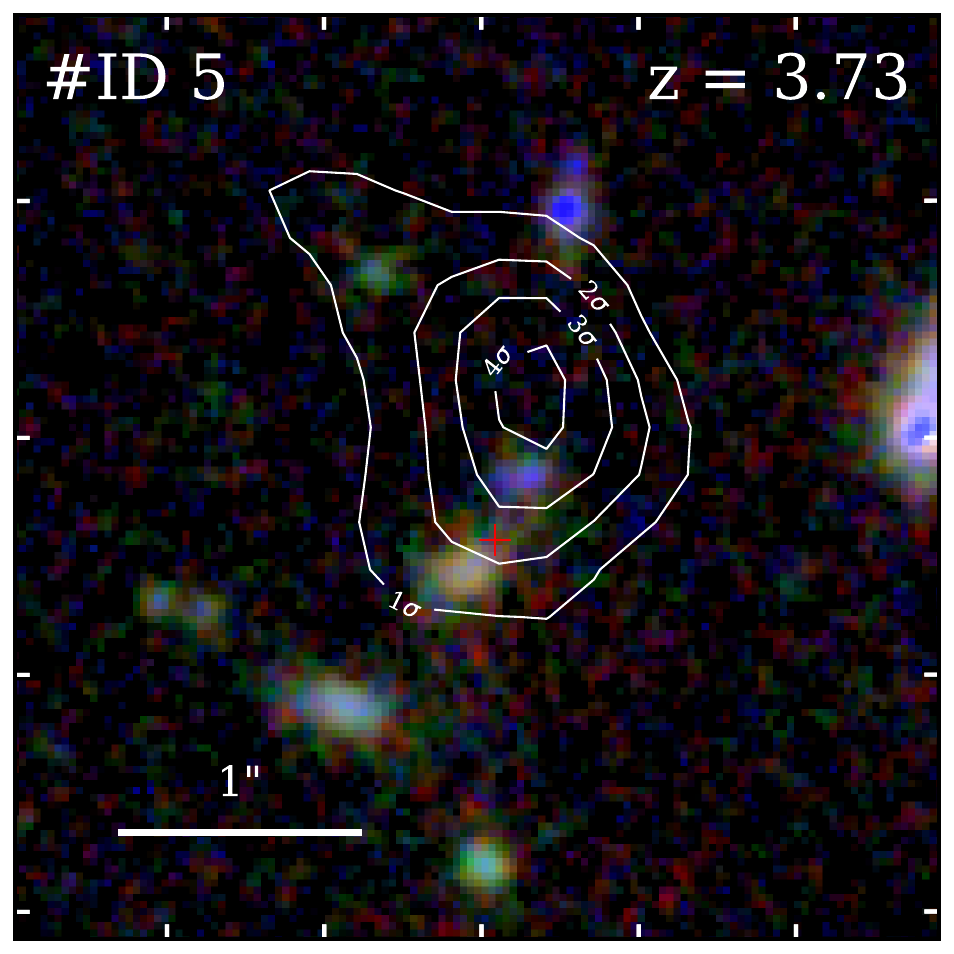}}
\hfill
\subfigure{\includegraphics[width = 2.3in, clip, trim = 0cm -0.5cm 0cm 0cm]{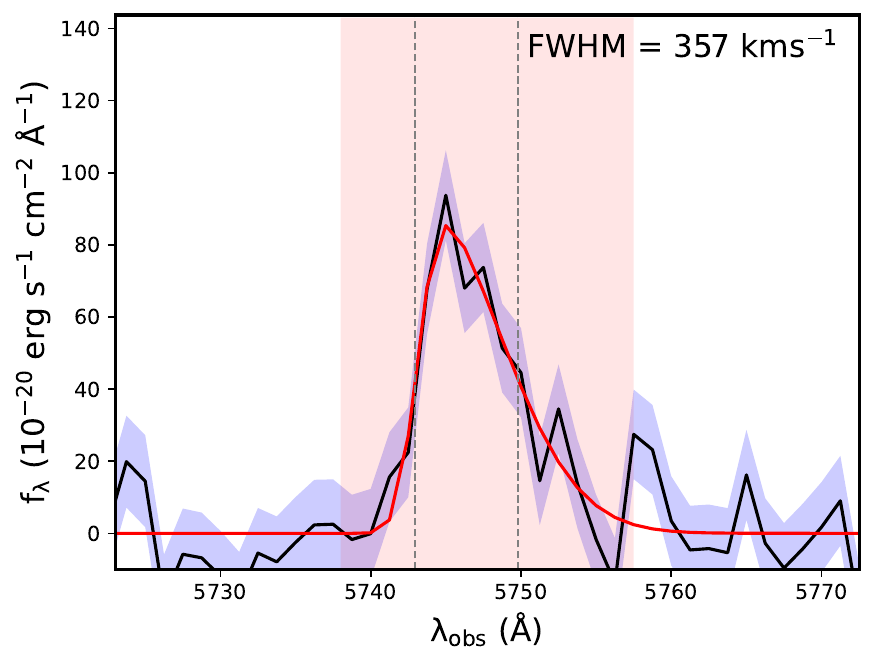}}
\hfill
\subfigure{\includegraphics[ width = 3in]{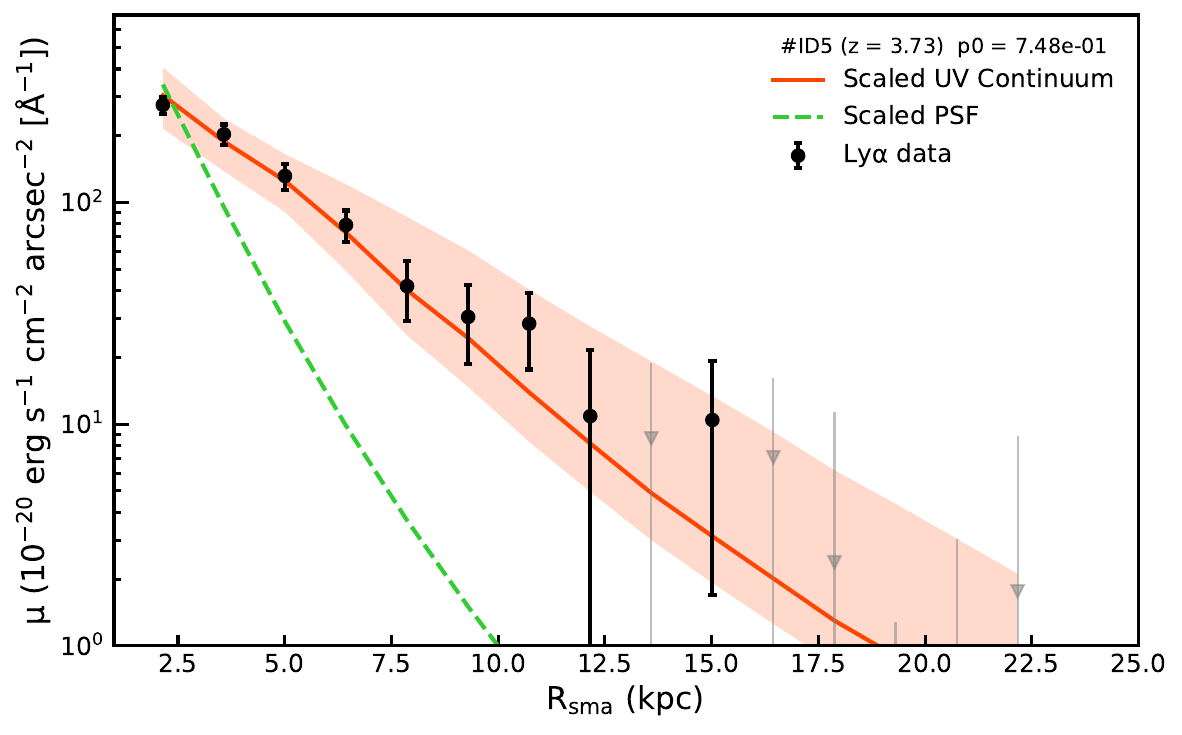}}\\

\vspace{-0.5cm}

\subfigure{\includegraphics[width = 1.67in, clip, trim = 0cm -2cm 0cm 0cm]{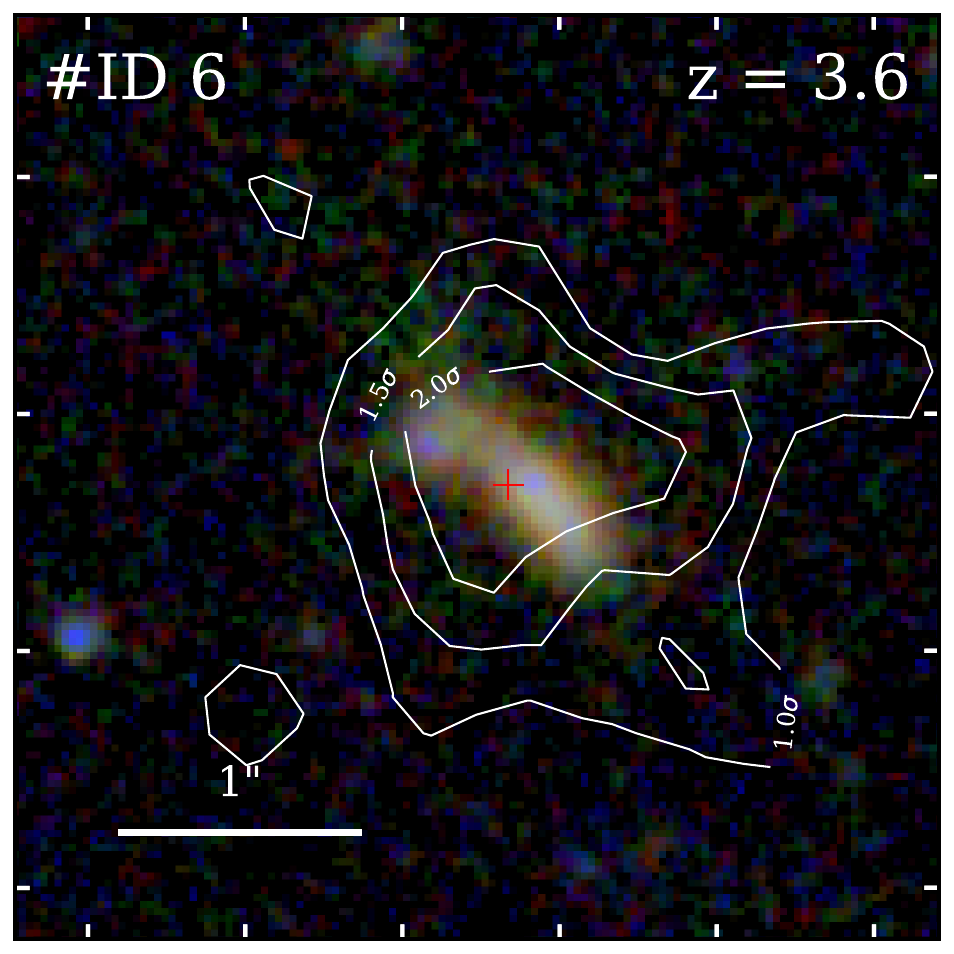}}
\hfill
\subfigure{\includegraphics[width = 2.3in, clip, trim = 0cm -0.5cm 0cm 0cm]{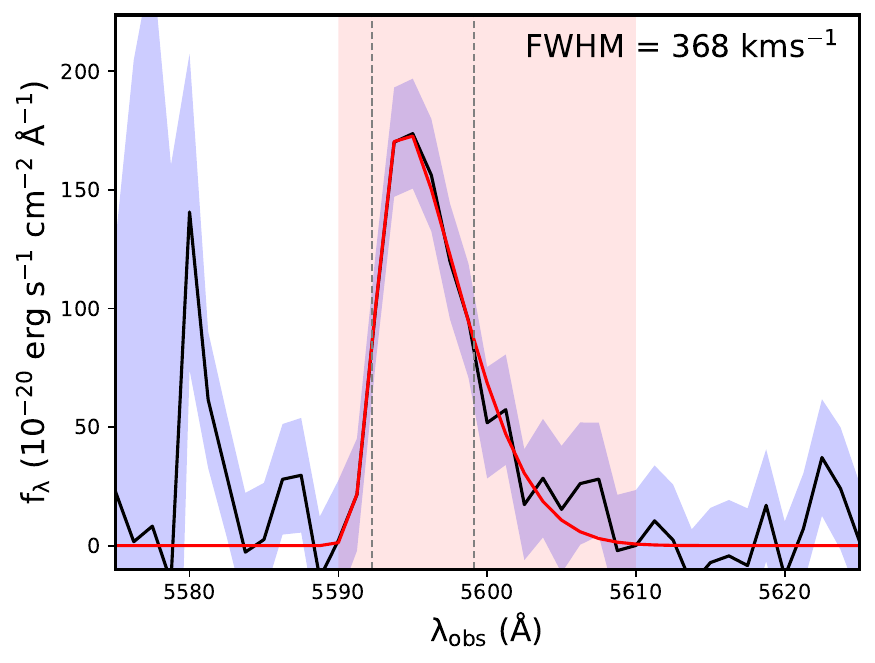}}
\hfill
\subfigure{\includegraphics[ width = 3in]{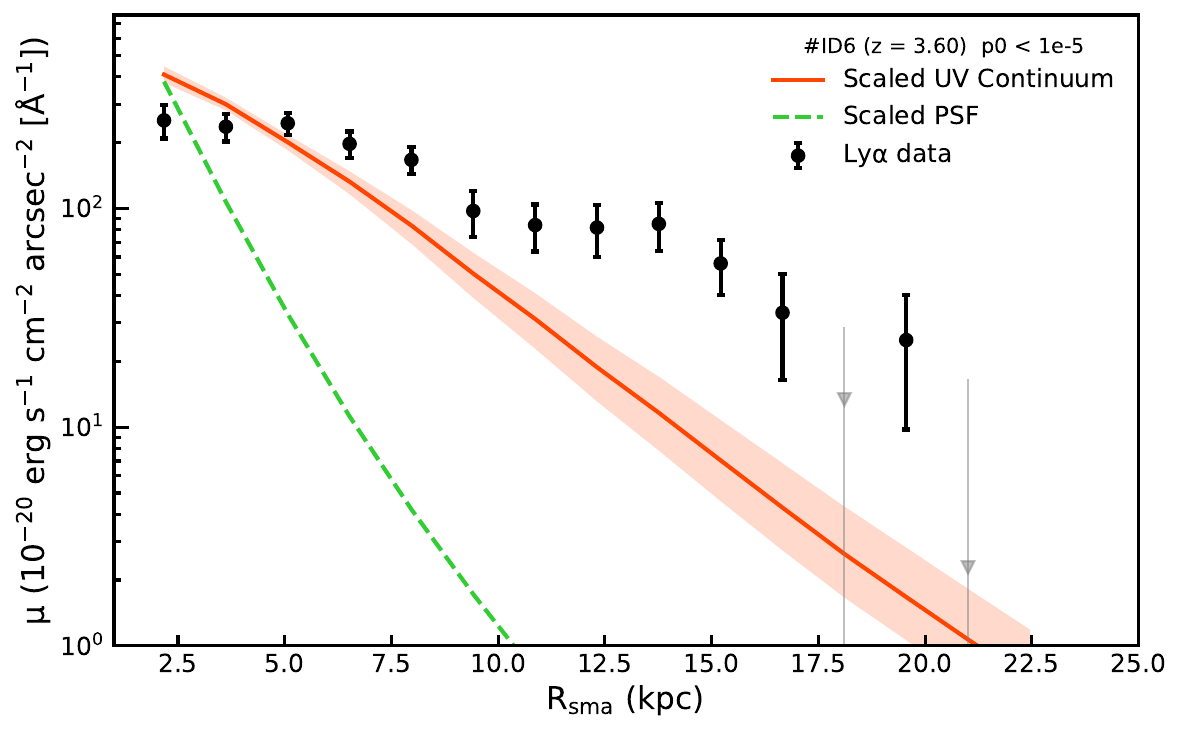}}\\

\vspace{-0.5cm}

\subfigure{\includegraphics[width = 1.67in, clip, trim = 0cm -2cm 0cm 0cm]{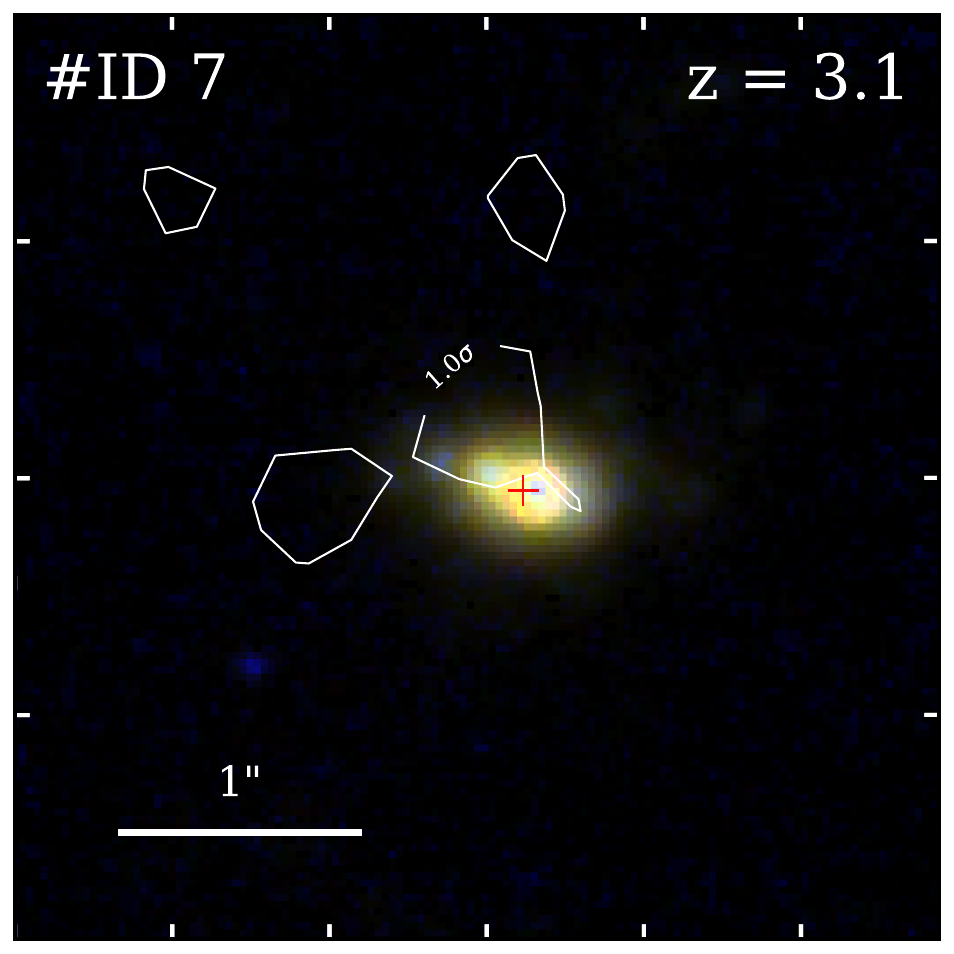}}
\hfill
\subfigure{\includegraphics[width = 2.3in, clip, trim = 0cm -0.5cm 0cm 0cm]{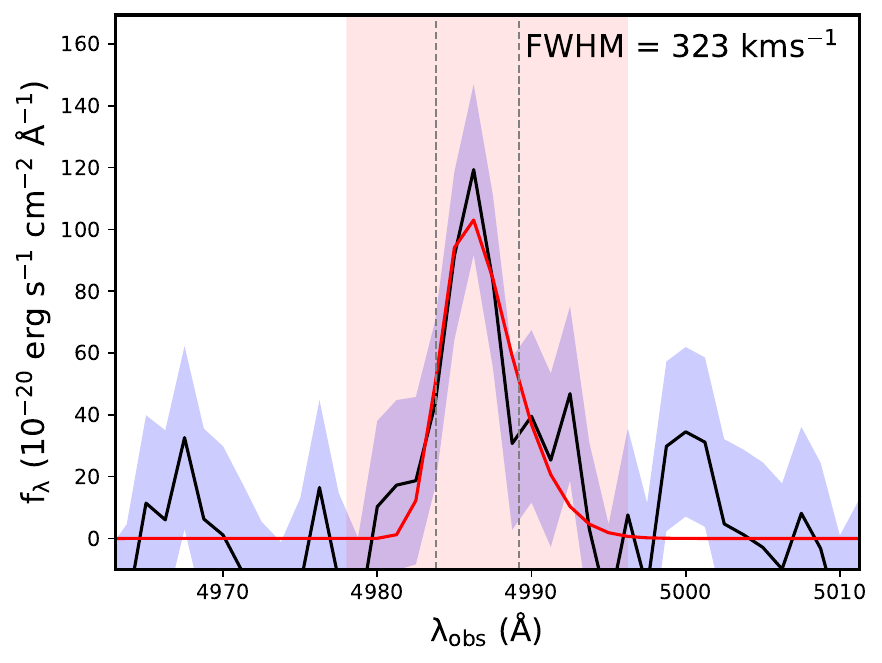}}
\hfill
\subfigure{\includegraphics[ width = 3in]{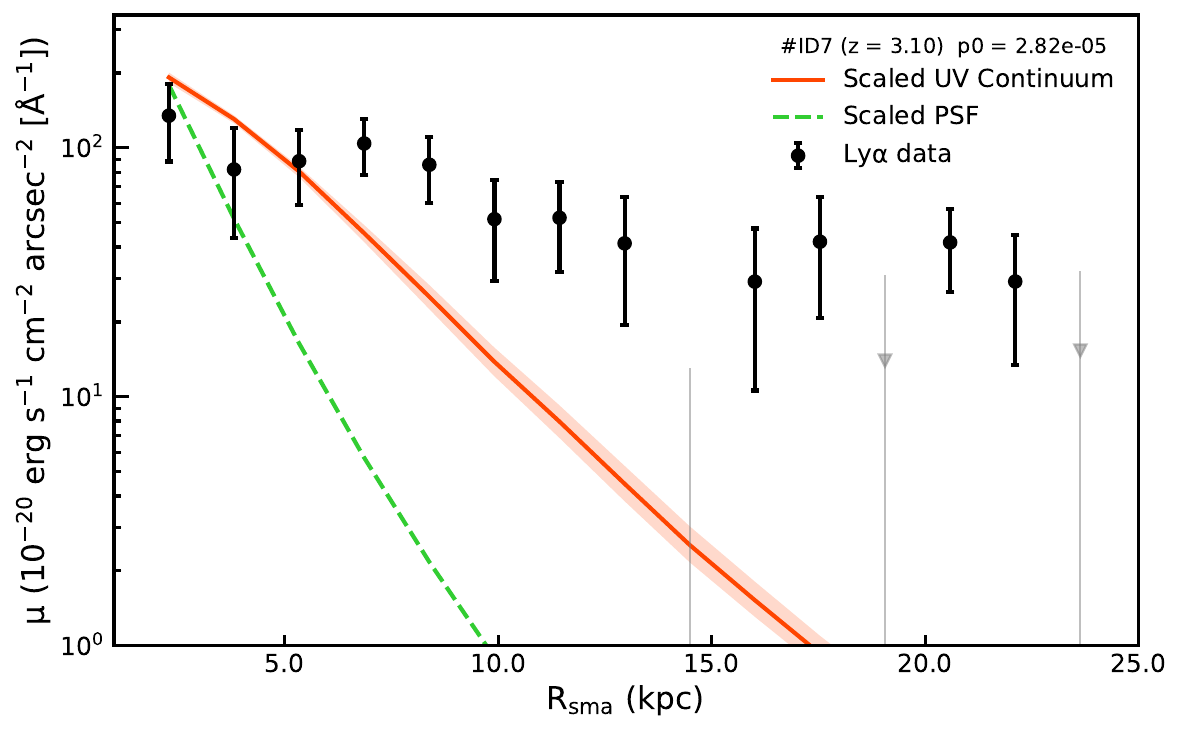}}\\
\caption{continuation of Figure~\ref{fig:Lya_halo_spec_rgb}}
\end{figure*}

\begin{figure*}[ht!]
\centering
\addtocounter{figure}{-1}

\subfigure{\includegraphics[width = 1.67in, clip, trim = 0cm -2cm 0cm 0cm]{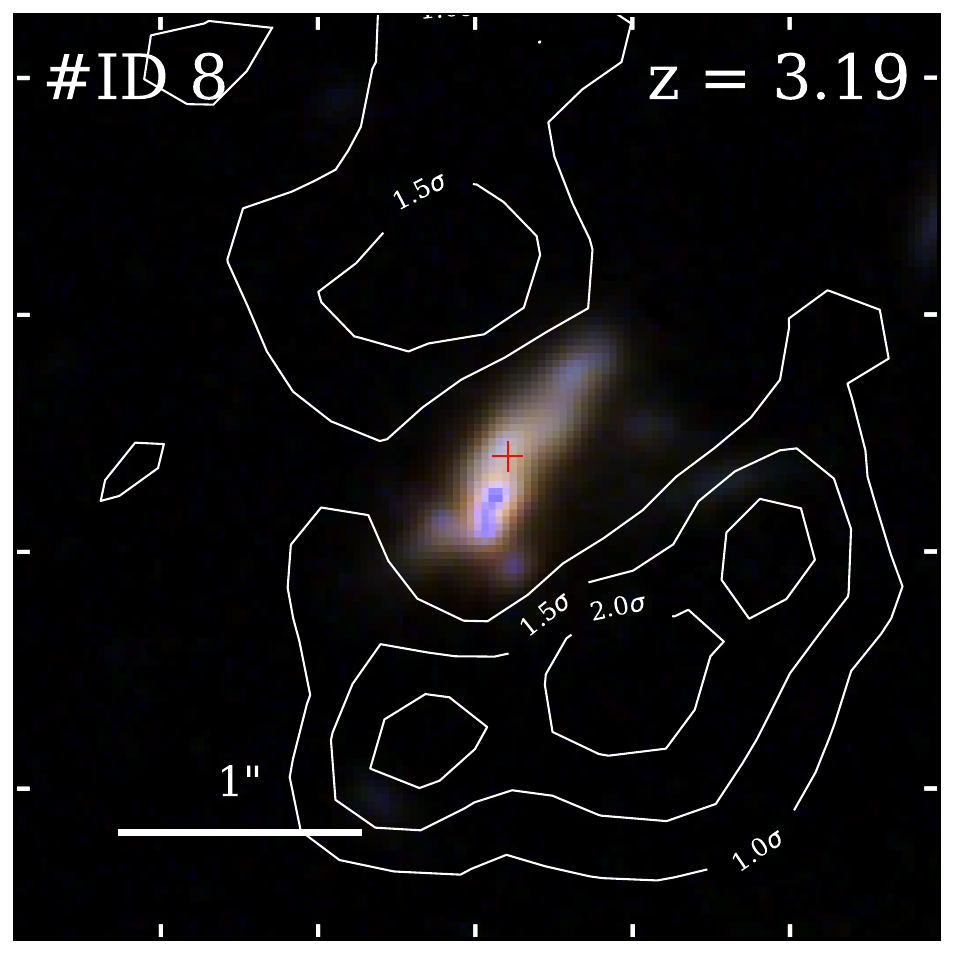}}
\hfill
\subfigure{\includegraphics[width = 2.3in, clip, trim = 0cm -0.5cm 0cm 0cm]{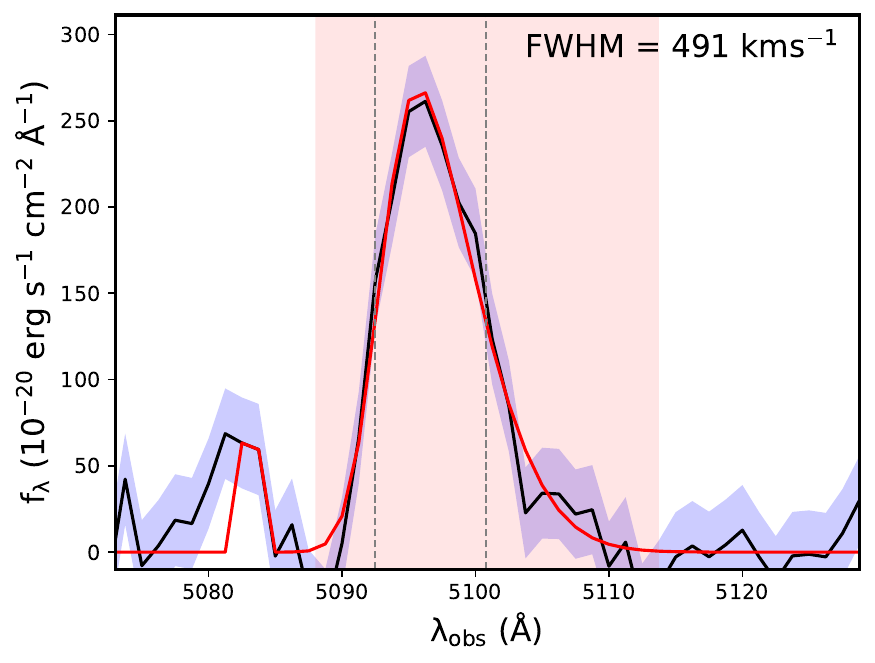}}
\hfill
\subfigure{\includegraphics[ width = 3in]{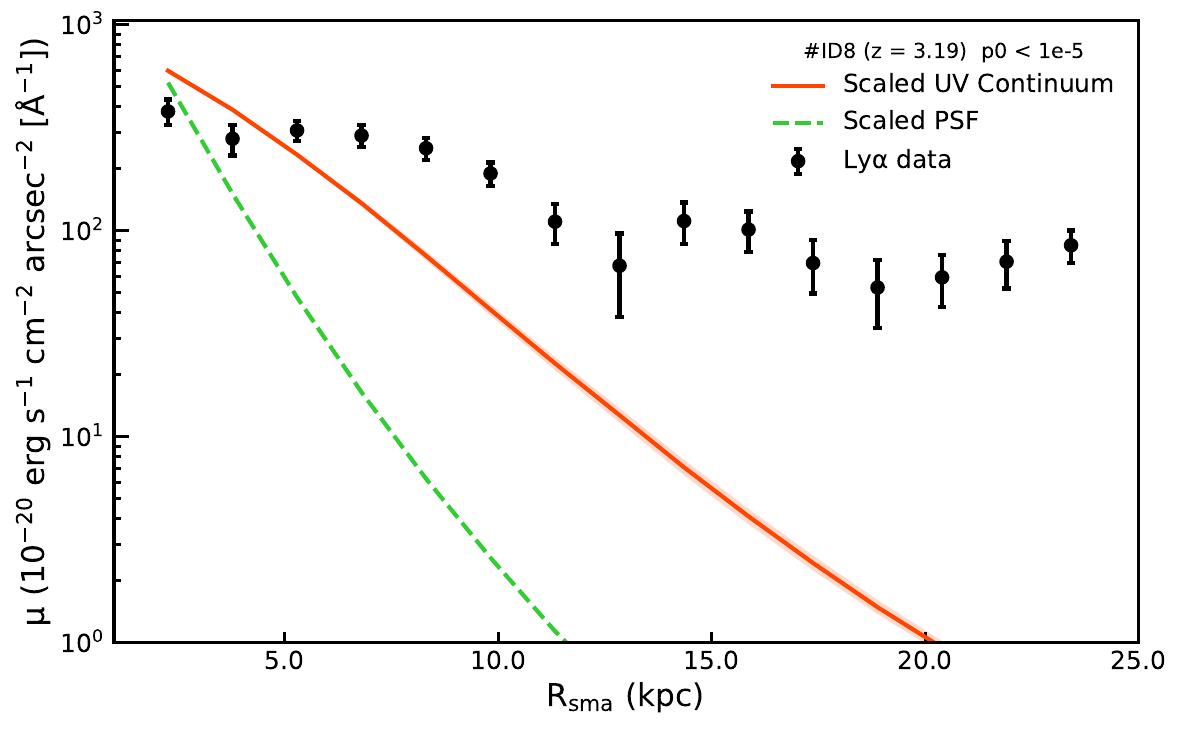}}\\

\vspace{-0.5cm}

\subfigure{\includegraphics[width = 1.67in, clip, trim = 0cm -2cm 0cm 0cm]{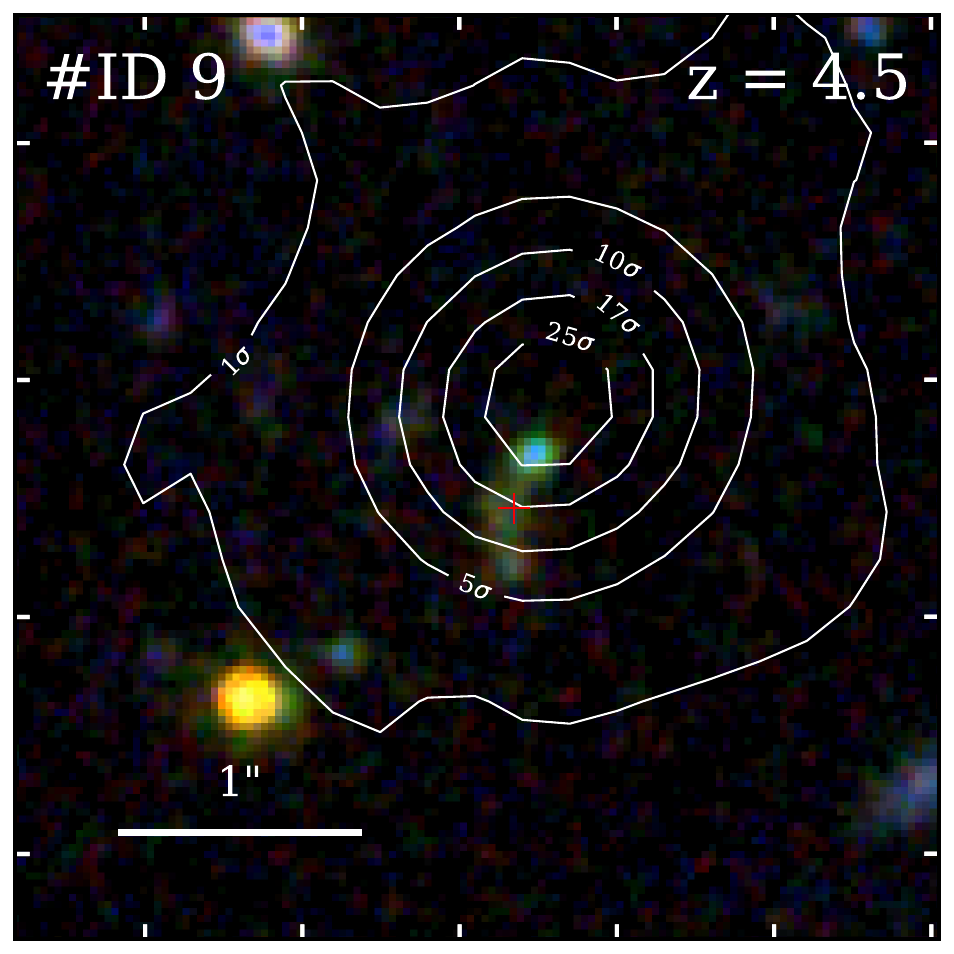}}
\hfill
\subfigure{\includegraphics[width = 2.3in, clip, trim = 0cm -0.5cm 0cm 0cm]{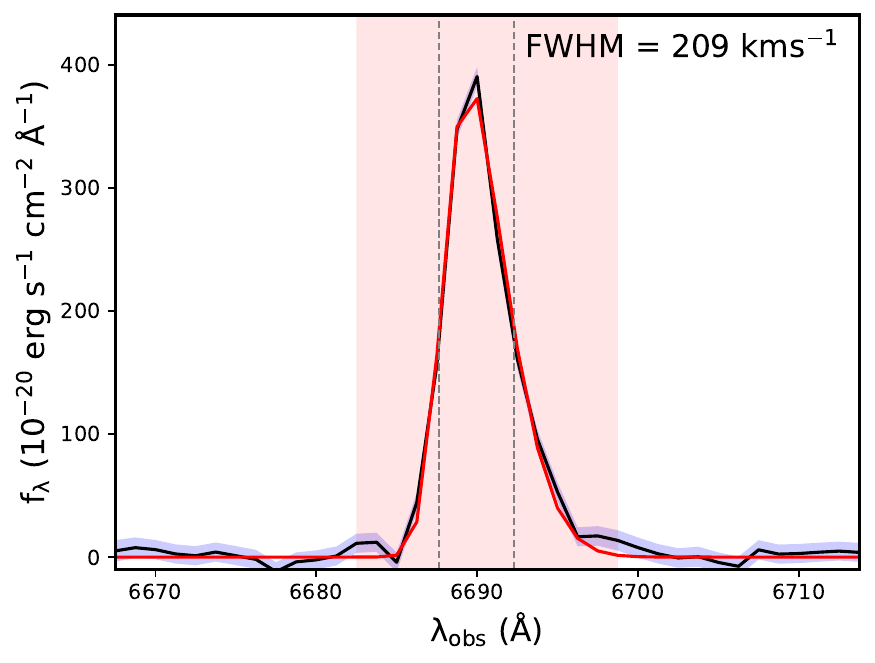}}
\hfill
\subfigure{\includegraphics[ width = 3in]{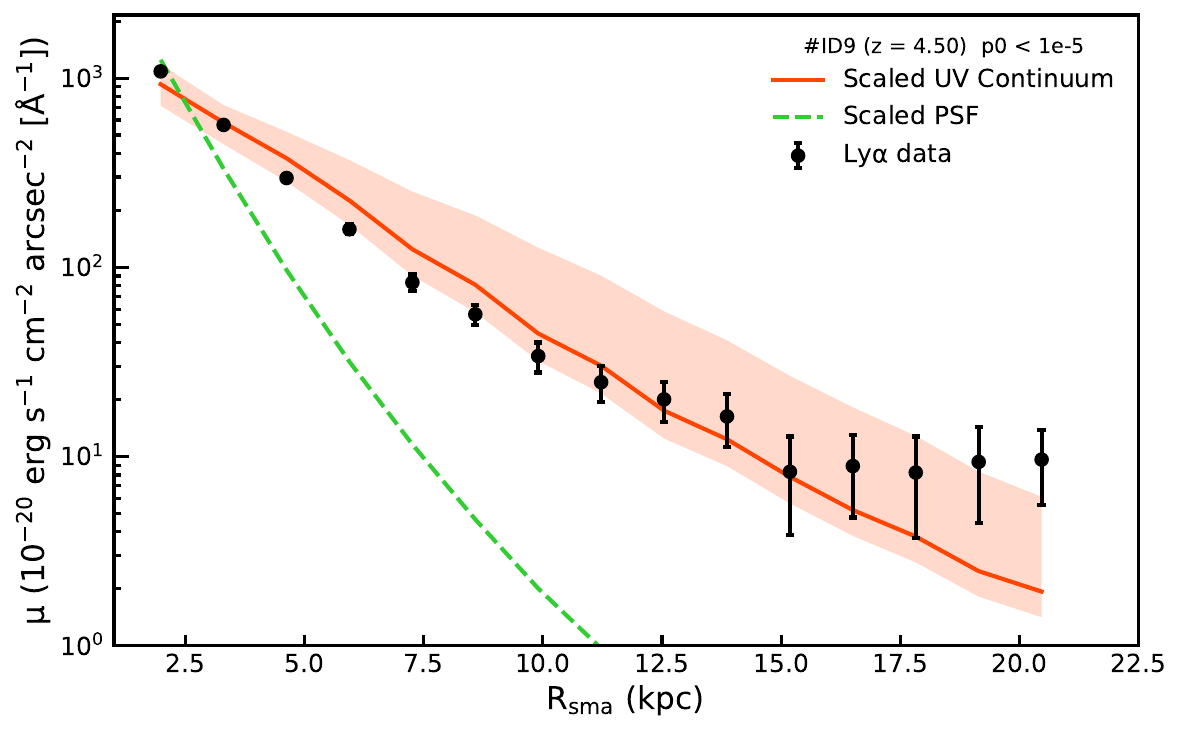}}\\

\vspace{-0.5cm}

\subfigure{\includegraphics[width = 1.67in, clip, trim = 0cm -2cm 0cm 0cm]{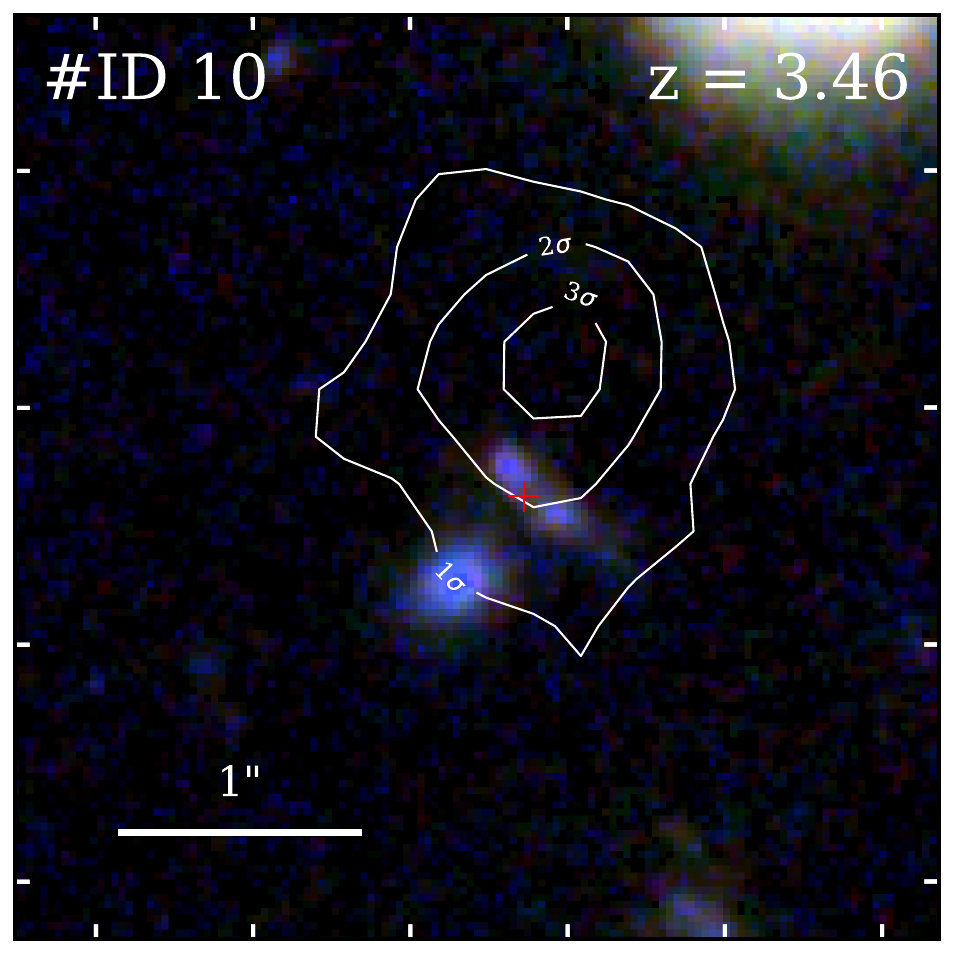}}
\hfill
\subfigure{\includegraphics[width = 2.3in, clip, trim = 0cm -0.5cm 0cm 0cm]{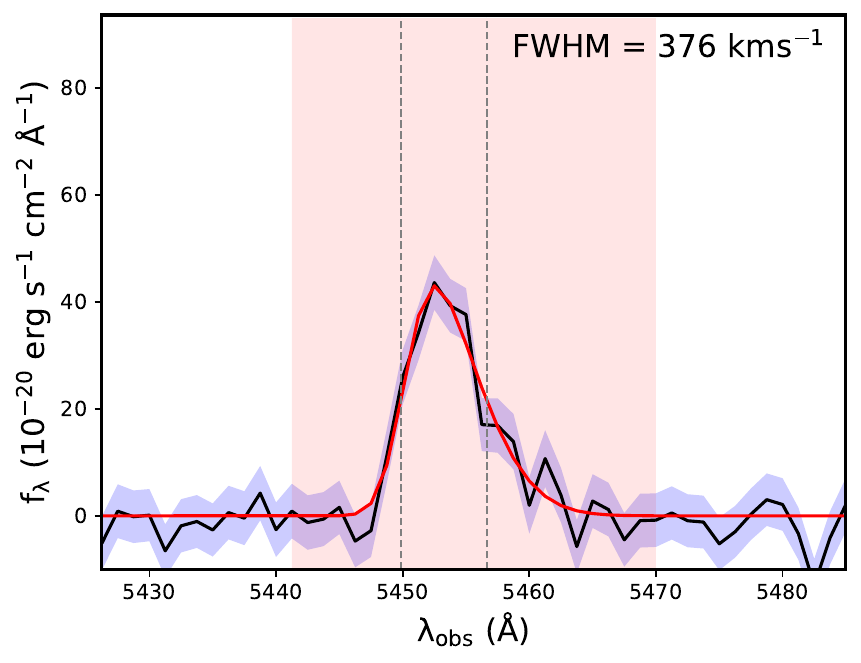}}
\hfill
\subfigure{\includegraphics[ width = 3in]{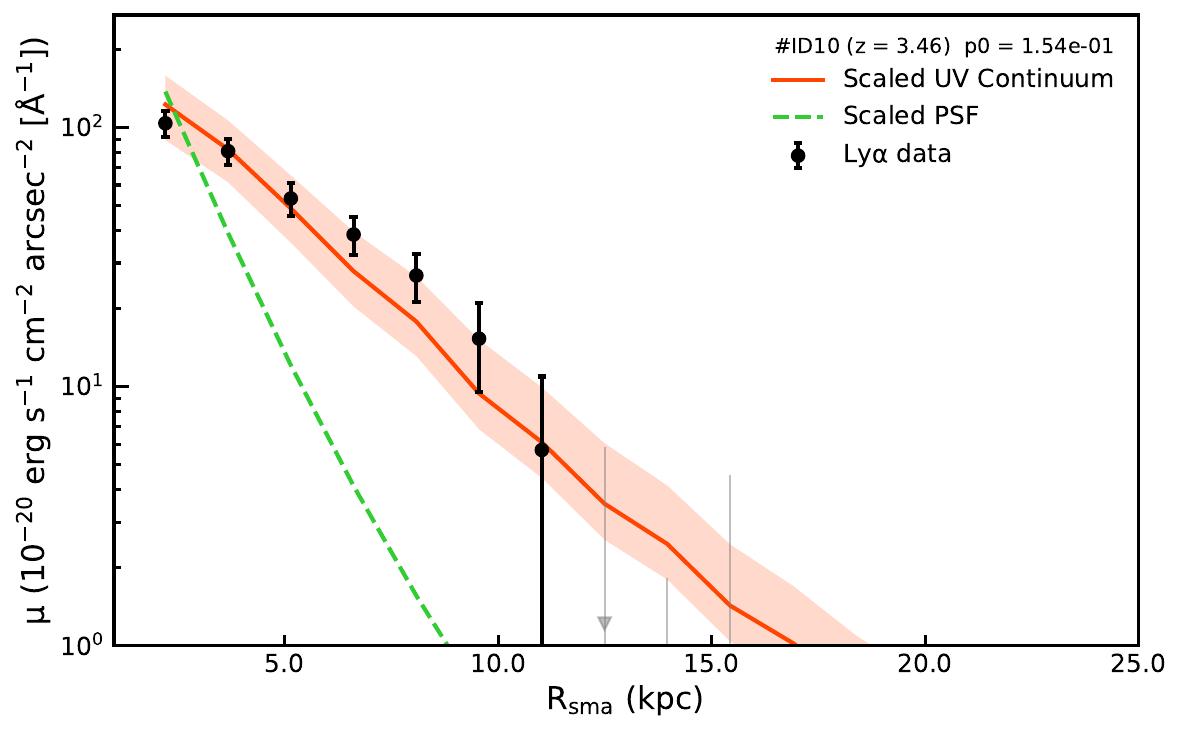}}\\

\vspace{-0.5cm}

\subfigure{\includegraphics[width = 1.67in, clip, trim = 0cm -2cm 0cm 0cm]{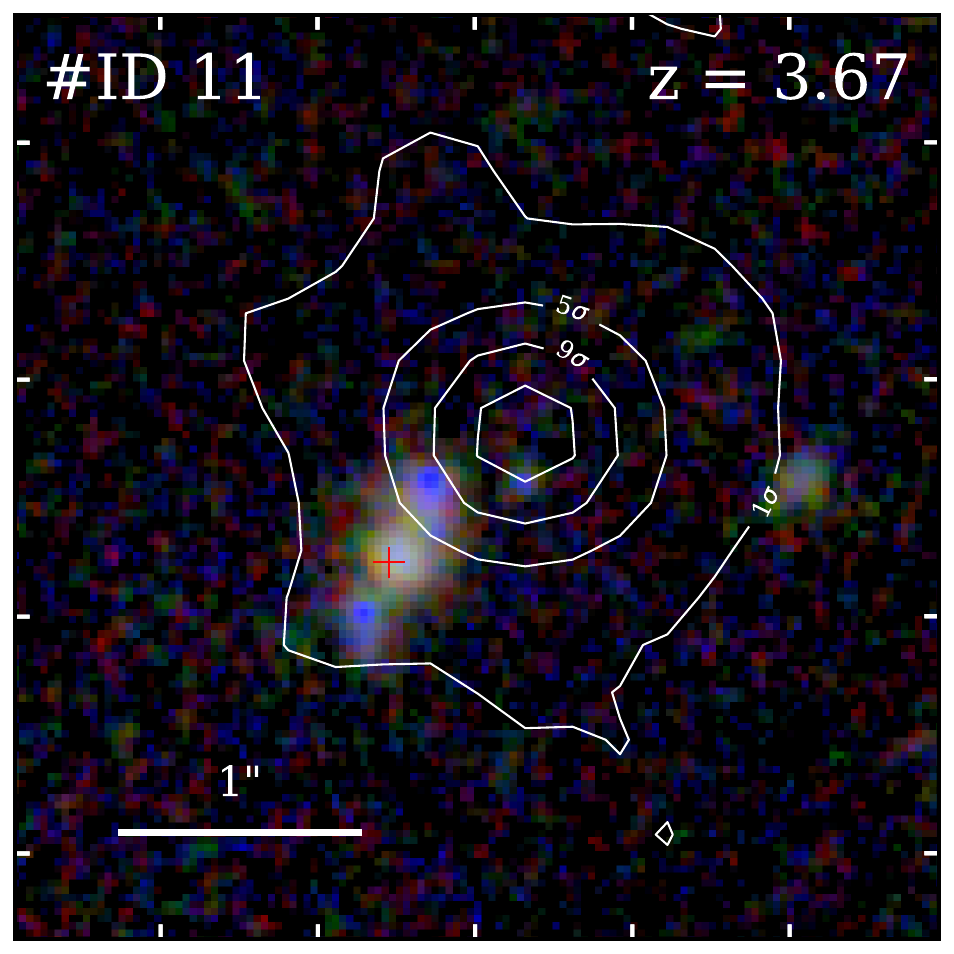}}
\hfill
\subfigure{\includegraphics[width = 2.3in, clip, trim = 0cm -0.5cm 0cm 0cm]{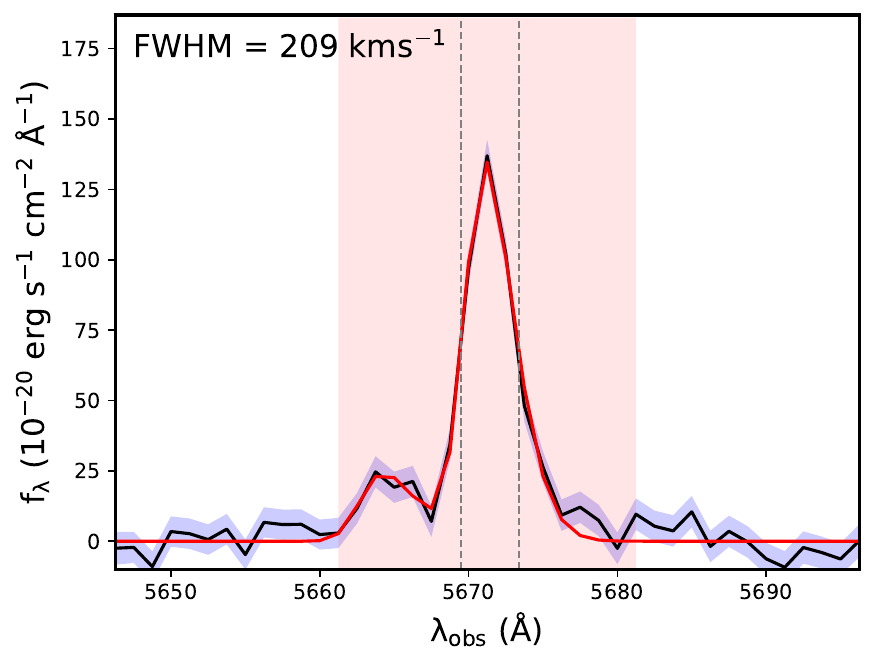}}
\hfill
\subfigure{\includegraphics[ width = 3in]{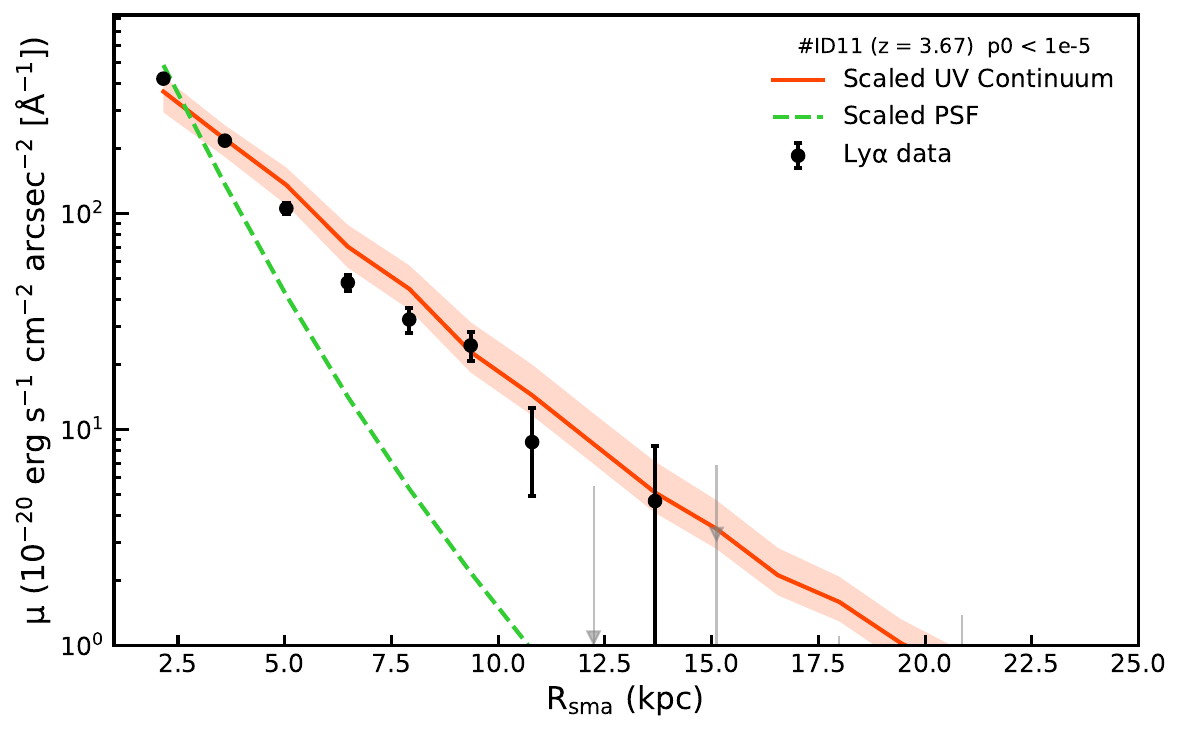}}\\

\vspace{-0.5cm}

\subfigure{\includegraphics[width = 1.67in, clip, trim = 0cm -2cm 0cm 0cm]{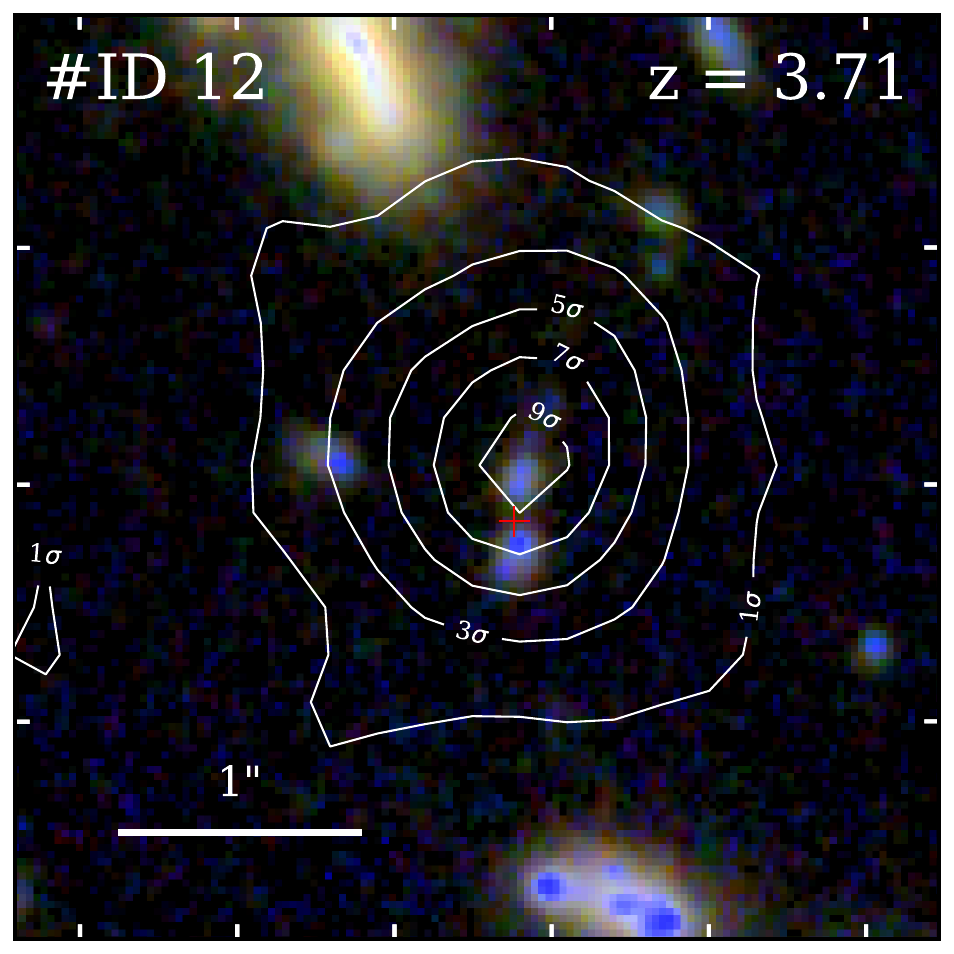}}
\hfill
\subfigure{\includegraphics[width = 2.3in, clip, trim = 0cm -0.5cm 0cm 0cm]{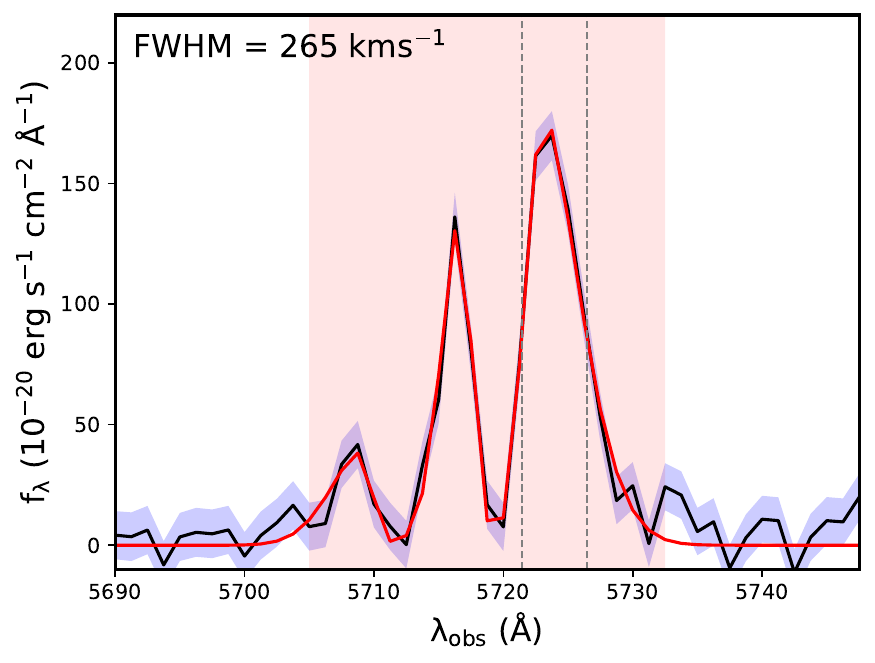}}
\hfill
\subfigure{\includegraphics[ width = 3in]{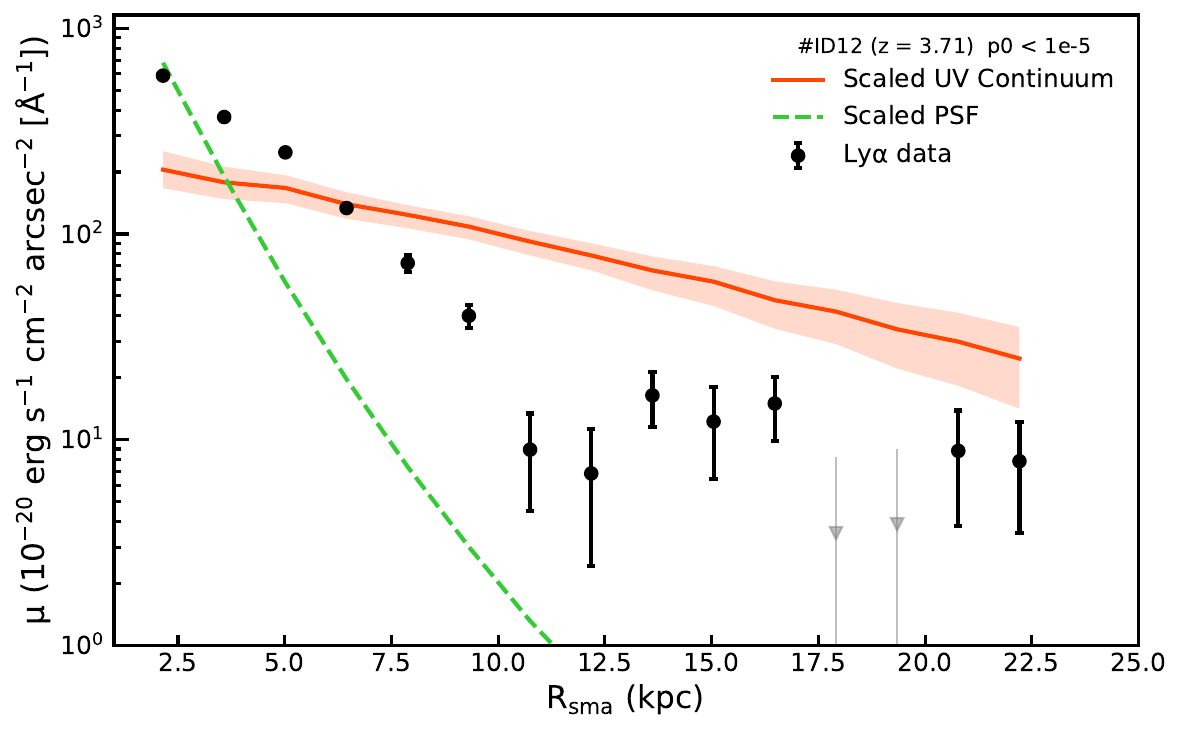}}\\
\caption{continuation of Figure~\ref{fig:Lya_halo_spec_rgb}}
\end{figure*}

After preparation of the $\rm Ly\alpha$ narrowband images and corresponding PSF-matched UV continuum images (see sec.\ref{sec:lya_nb_img} and \ref{sec:uvc_img}), we measure the surface brightness profile of these emission maps. We use the \texttt{find\_galaxy()} method from the Python module \texttt{mgefit} \citep{Cappellari2002} to estimate the center, PA, and ellipticity from the images. This method uses the weighted first and second moments of the intensity to compute these parameters, assuming that the emission from the target galaxy is the largest cluster of positive values in the images. 

We first identify the parameters as mentioned above, and then we fix these geometrical parameters and put concentric elliptical annuli with a separation of 1~pixel ($\rm 0.2^{\prime\prime}$), and then calculate the intensity within each annulus and assign it to the radius, which is the center of the annulus. We do the same procedure to obtain the surface brightness profile of both the UV continuum and $\rm Ly\alpha$ narrowband images. 

\subsection{Error estimation for surface brightness profile} \label{sec:sb_prof_err}

\begin{figure}
    \centering
    \includegraphics[width=\linewidth]{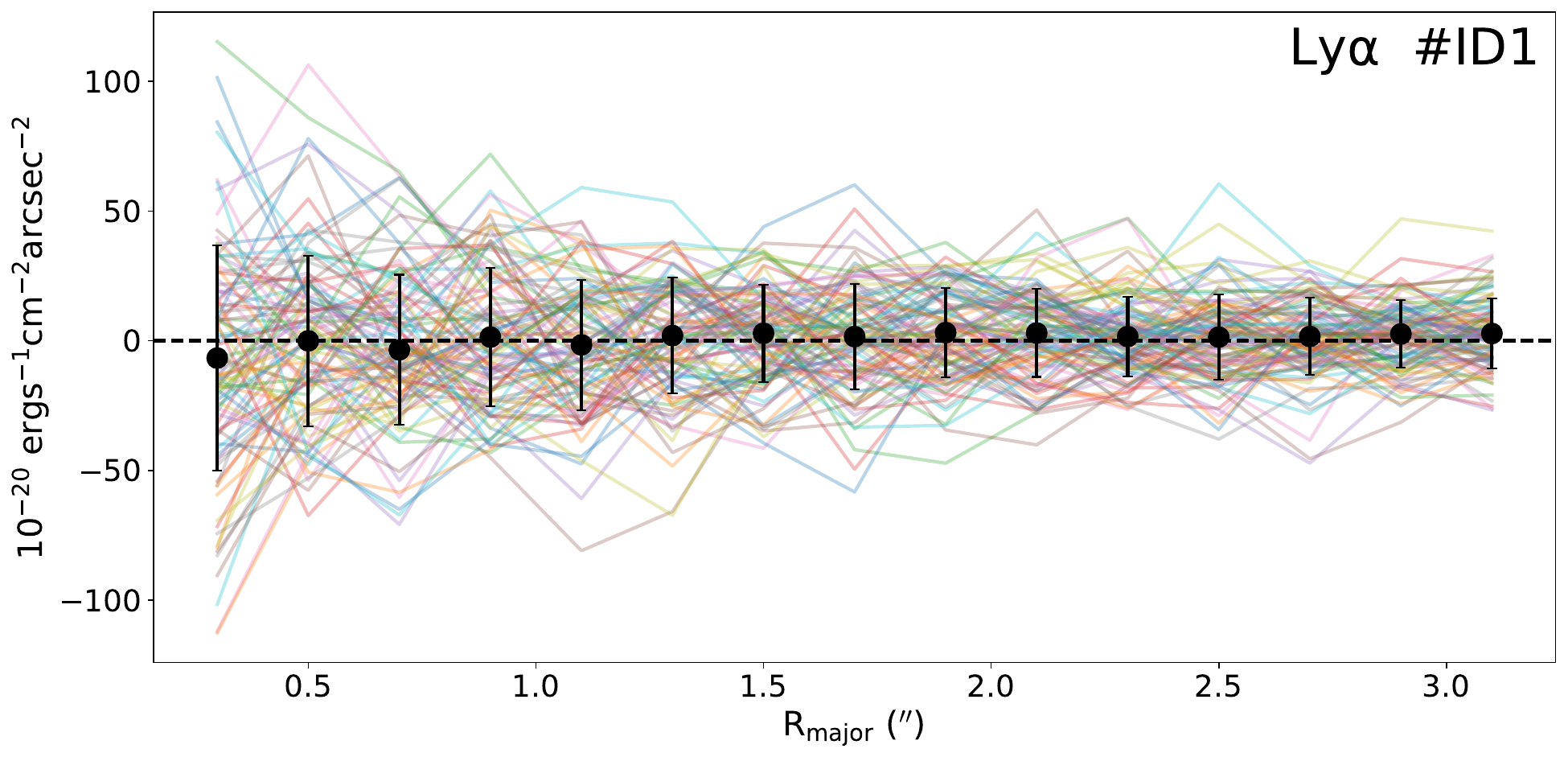} \\
    \includegraphics[width = \linewidth]{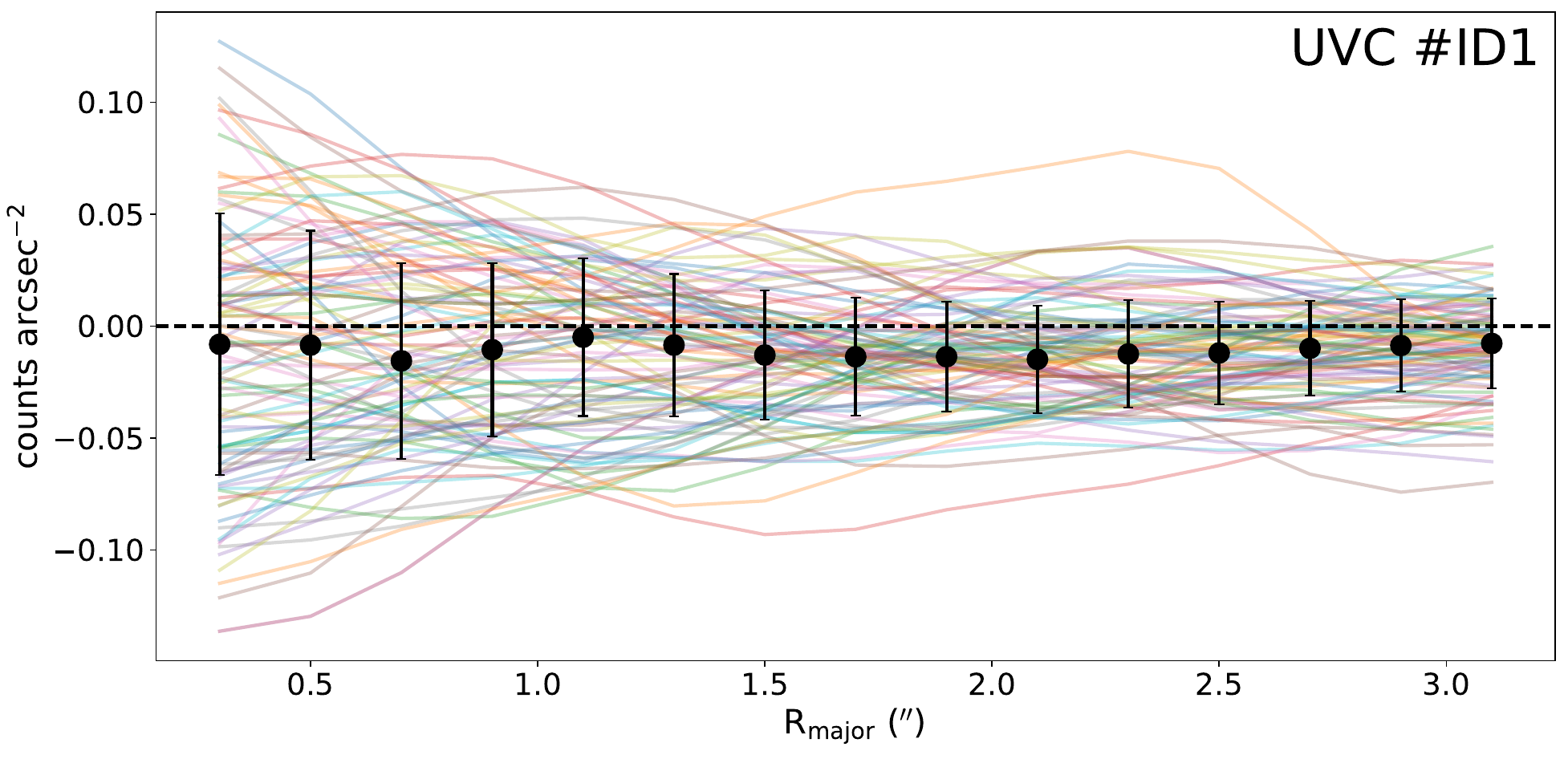}
    \caption{Plot shows the radial profile of the empty background patches taken from the narrow band imaging in the case of the top panel ($\rm Ly\alpha$) and the UV continuum band image using the same elliptical annulus as used for the galaxy. The black points show the median of all 100 points at a given radius, and the error bars are taken as the standard deviation of the points.}
    \label{fig:uv_lya_err_plot}
\end{figure}

To compute the error in the surface brightness values, we adopt a method similar to that of \cite{Wisotzki_etal2016}. For the MUSE $\rm Ly\alpha$ surface brightness profiles errors, we construct the narrow band image of the whole field by collapsing in the same wavelength range as the narrow band for each object (see the middle column of Fig.~\ref{fig:Lya_halo_spec_rgb}). We then generate a segmentation map for each image, select 100 empty regions, and save cutouts of these regions, of the same shape as the narrowband images. The same annulus has been placed over the empty regions used to measure the surface brightness profile. After performing the same procedure on 100 images, we take the median and the standard deviation, the latter of which has been used as an error value for the datapoints. The median helps detect residual background in the narrow-band images (see Figure~\ref{fig:uv_lya_err_plot}). 

We follow the same procedure for the UV images and obtain HST empty-region cutouts. These cutouts are then rescaled to the MUSE pixel scale; are convolved with the respective MUSE PSF, and place the annulus.

In Fig.~\ref{fig:uv_lya_err_plot} we show the profiles of the 100 empty patches in $\rm Ly\alpha$ and UV emission. We can see that both profiles show a large variation in the smaller-radius range, then fall off sharply with increasing radius. The UV error profiles look smoother in comparison to the $\rm Ly\alpha$ due to convolution with the MUSE PSF; the noise peaks also smooth out, which in turn smooths the noise features in the original image. In both plots, the median shown with the black dot appears consistent with no residual background, and the error bars represent the standard deviation across 100 empty patches.

\section{Flux and equivalent width of \texorpdfstring{$\rm Ly\alpha$\ }\ emission} \label{sec:flux_and_ew_lya}

To measure the total $\rm Ly\alpha$ flux, we employed a curve-of-growth method. For each galaxy, we first estimate the background noise by measuring the standard deviation in empty regions of the full field $\rm Ly\alpha$ narrowband images. A surface brightness threshold corresponding to $\rm 3\sigma$ above the background was then adopted. Surface brightness profiles were constructed using circular annuli centered on the galaxy position determined with the \texttt{find\_galaxy()} module (see Section~\ref{sec:sb_profile}). The aperture radius was defined as the point where the surface brightness profile reached the adopted threshold, and the total $\rm Ly\alpha$ flux was measured within this aperture. 

\begin{equation}
EW_{0} = \frac{1}{(1+z)}\frac{F_{Ly\alpha}}{f_{\lambda,UV}} \label{eqn:lya_ew}
\end{equation}

To calculate the rest-frame equivalent width of the $\rm Ly\alpha$ emission, we follow the procedure described in \cite{Drake_etal2017}. The HST filters were selected based on the optimal band choice for the corresponding redshift range following \cite{Hashimoto_etal2017}. Using three-band photometry sampling the rest-frame UV continuum between 1200--3000~\text{\AA}, we estimated the continuum flux density, $f_{\lambda,\mathrm{UV}}$, by fitting a straight line to the observed photometric measurements. The rest-frame equivalent width, $\rm EW_{0}$, was then calculated using Eqn.~\ref{eqn:lya_ew}.

\begin{figure*}[ht!]
\centering
\subfigure{\includegraphics[width = 3.5in, clip, trim = -0cm -0cm 0cm 0cm]{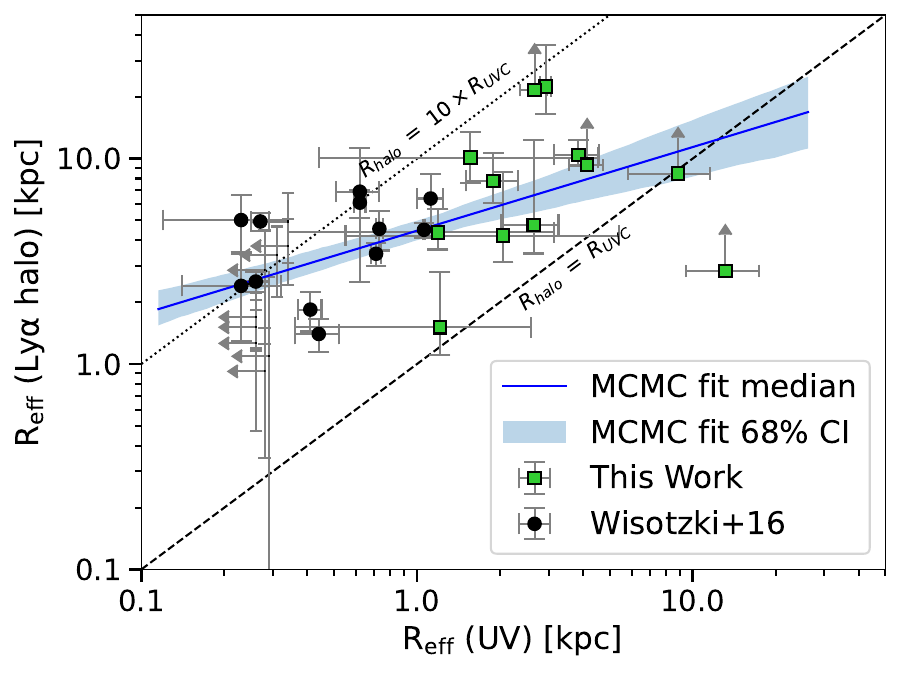}}
\hfill
\subfigure{\includegraphics[width = 3.5in, clip, trim = 0cm -0cm 0cm 0cm]{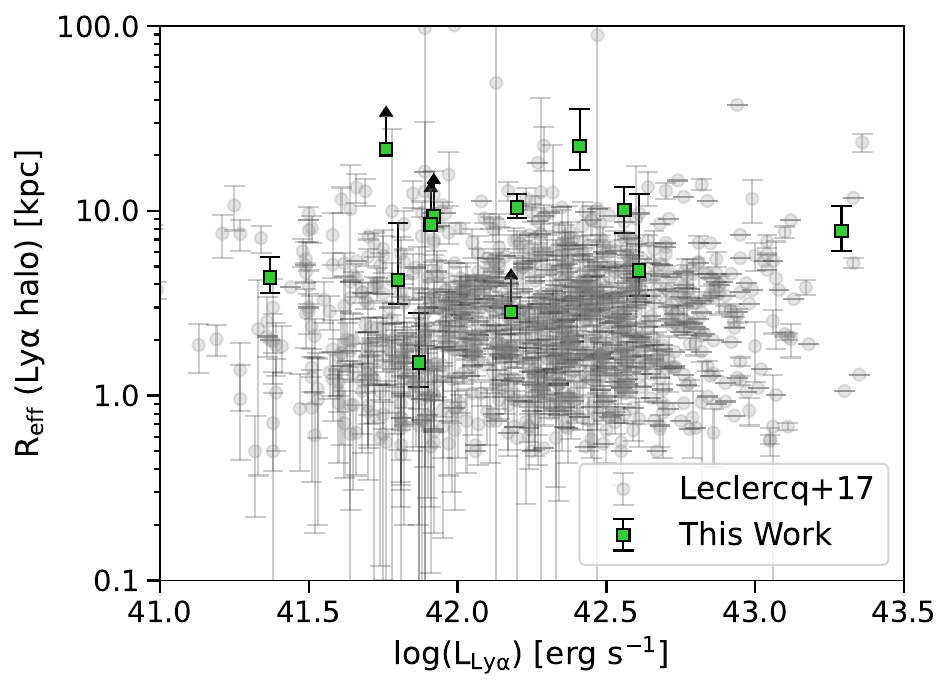}}\\

\vspace{-0.5cm}

\subfigure{\includegraphics[width = 3.5in, clip, trim = -0cm -0cm 0cm 0cm]{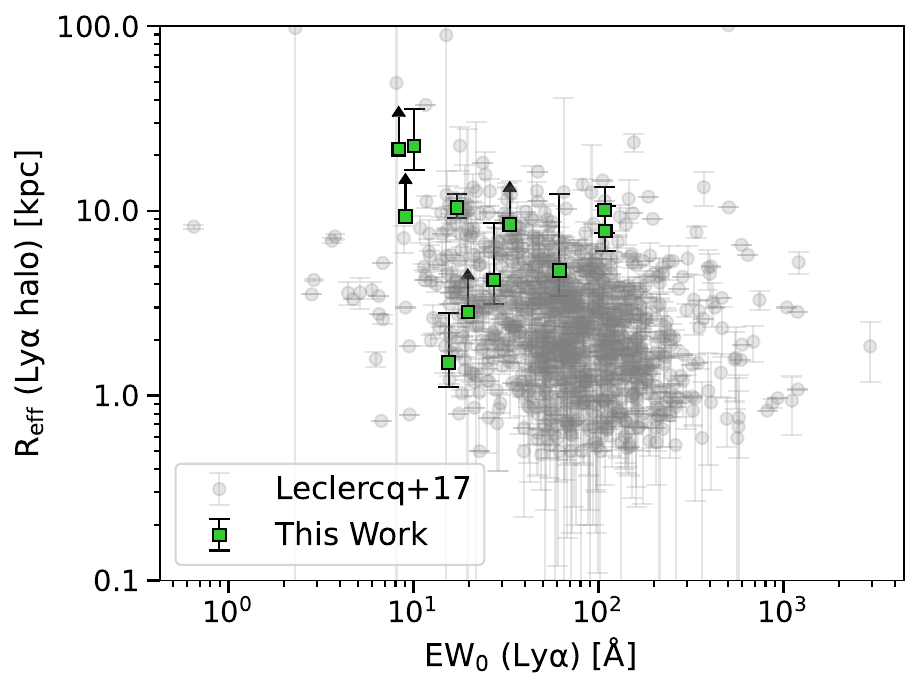}} 
\hfill
\subfigure{\includegraphics[width = 3.5in, clip, trim = -0cm -0cm 0cm 0cm]{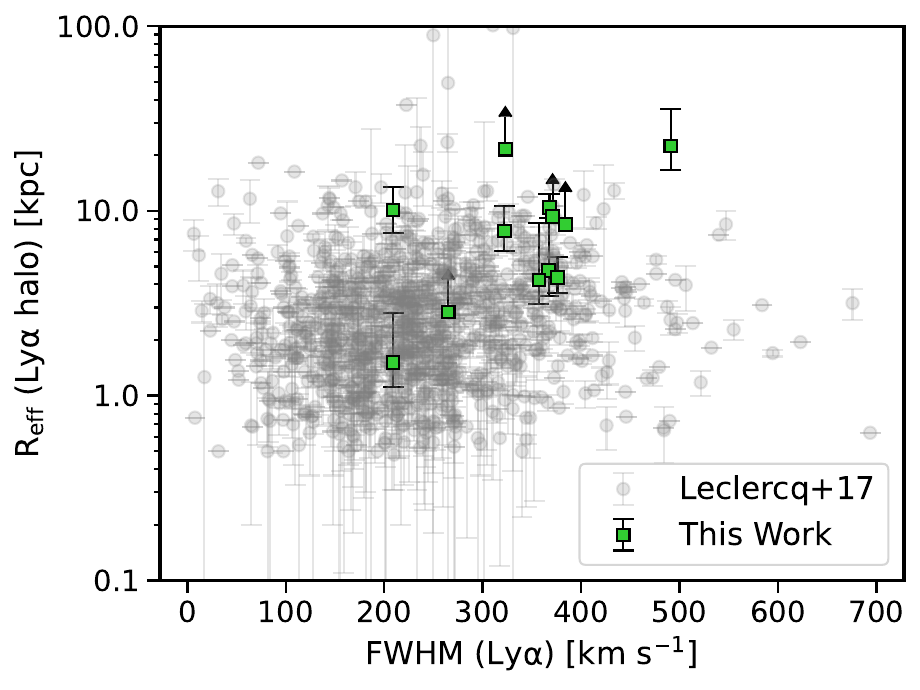}}

\caption{Correlation of $\rm Ly\alpha$ halo scale length with different parameters. The top-left plot shows the UV continuum scale length relative to the $\rm Ly\alpha$ halo scale length; the black circles show the measurements, while the green squares represent the measurements from the sample in this work. The dashed line shows the relation with the UV scale length equal to the $\rm Ly\alpha$ halo scale length; the other line shows $\rm 10\times$ the UV scale length; the blue line with a sky-blue band represents an MCMC fit (see Eqn~\ref{eqn:lya_uv_fit}) to the data with the band showing 1$\sigma$ error on the fit (68\% CI). The top-right panel shows the $\rm Ly\alpha$ halo scale length as a function of log luminosity, with the sources in agreement with the scatter observed in previous studies. The bottom-left plot shows the halo scale length comparison with the rest equivalent width ($\rm EW_{0}$). The bottom-right plot shows the correlation with the FWHM measured from the red part of the $\rm Ly\alpha$ emission-line fit, which, as before, does not deviate from previous trends.}

\label{fig:uv_lya_size_vs_prop}
\end{figure*}

Although a weak anti-correlation between halo scale length and $\rm Ly\alpha$ luminosity was reported by \cite{Leclercq_etal2017}, we do not observe any statistically significant correlation within our sample from the visual inspection of the plot, as shown in Fig.~\ref{fig:uv_lya_size_vs_prop}. Similarly, we find no strong dependence of the halo scale length on either $\rm log(L_{Ly\alpha})$ or the $\rm Ly\alpha$ line FWHM. The absence of clear trends in our sample is broadly consistent with previous studies of high-redshift $\rm Ly\alpha$ emitters.

\section{UV and \texorpdfstring{$\rm Ly\alpha$\ }\ sizes} \label{sec:uv_lya_sizes}

\begin{figure*}[ht!]
\centering

\subfigure{\includegraphics[width = 2.3in, clip, trim = 0cm 0cm 0cm 0cm]{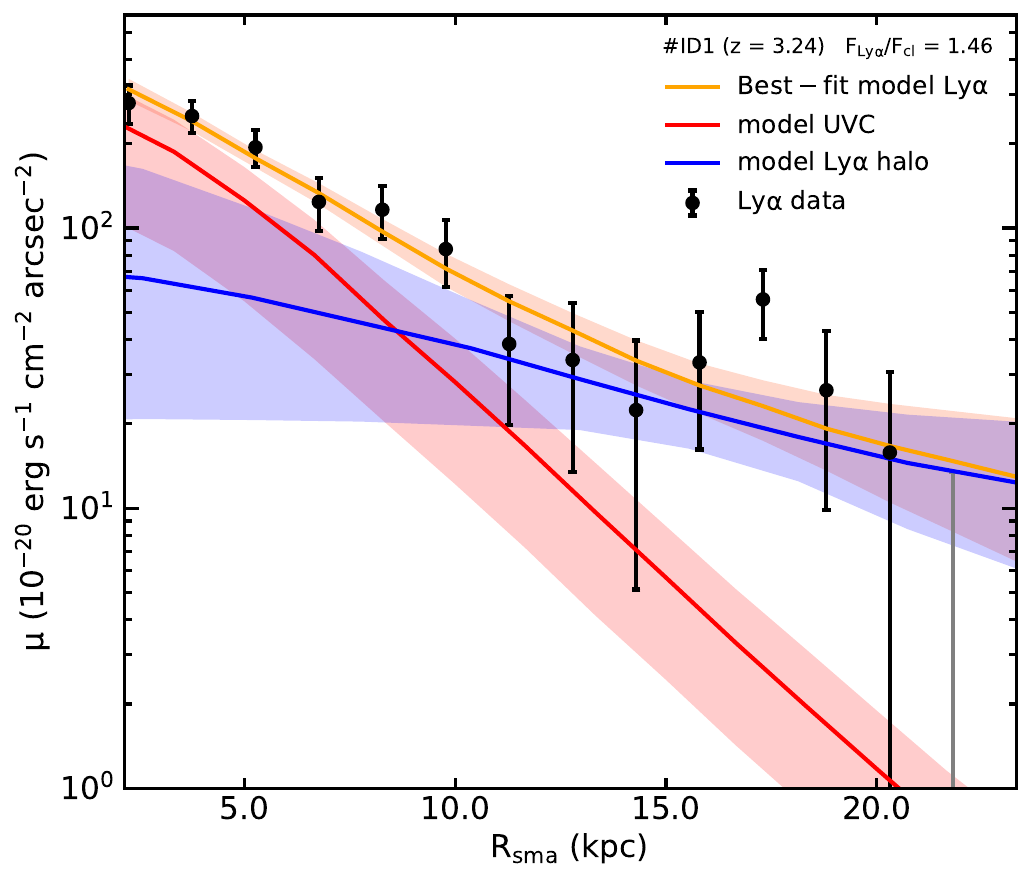}}
\hfill
\subfigure{\includegraphics[width = 2.3in, clip, trim = 0cm 0cm 0cm 0cm]{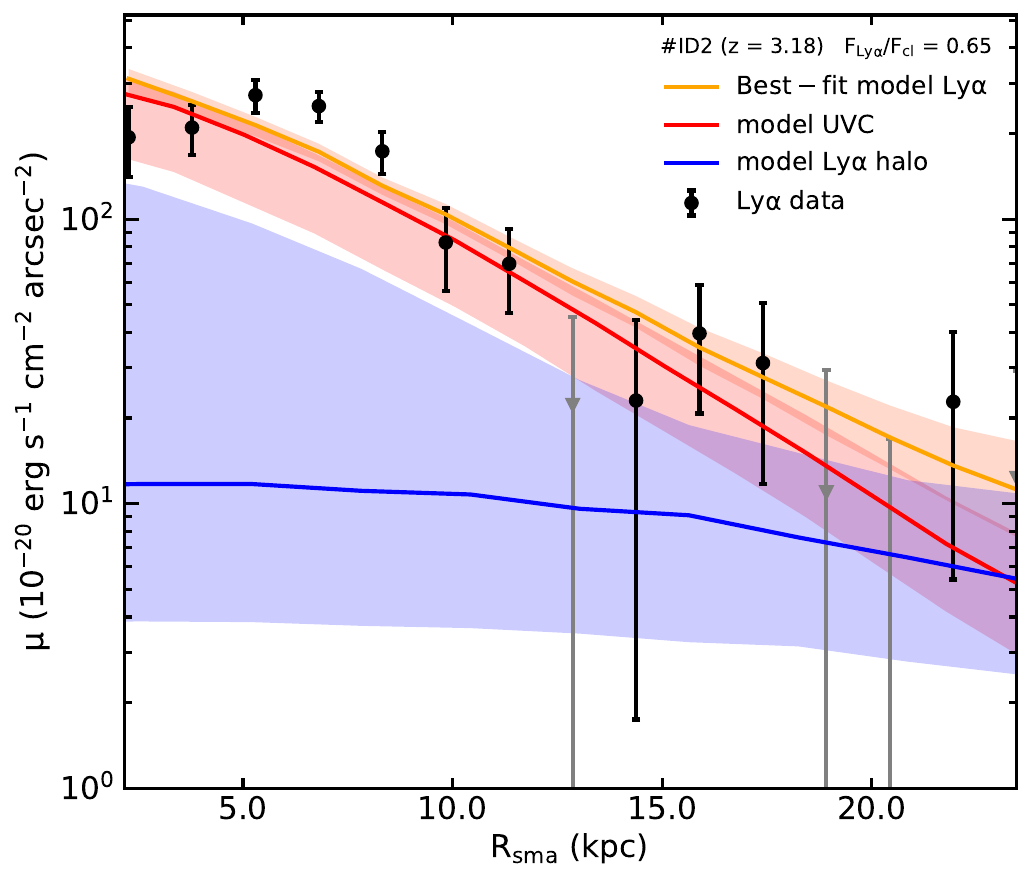}}
\hfill
\subfigure{\includegraphics[width = 2.3in]{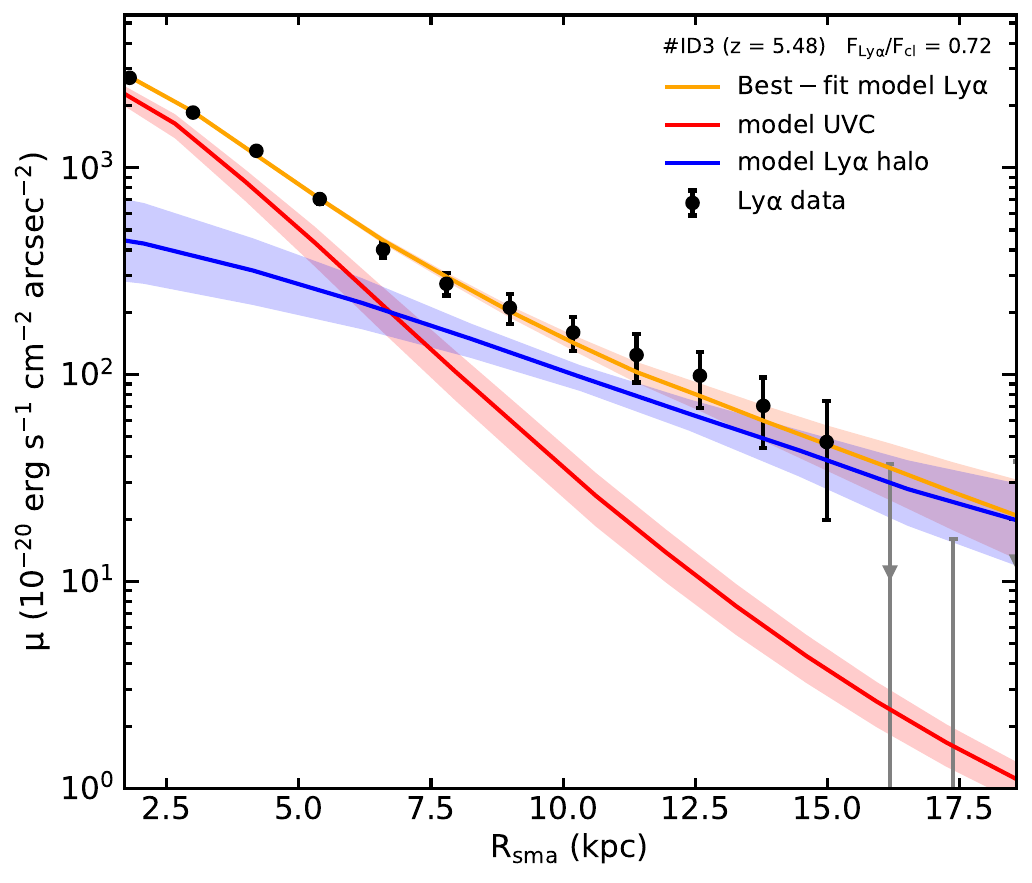}}\\

\subfigure{\includegraphics[width = 2.3in, clip, trim = 0cm 0cm 0cm 0cm]{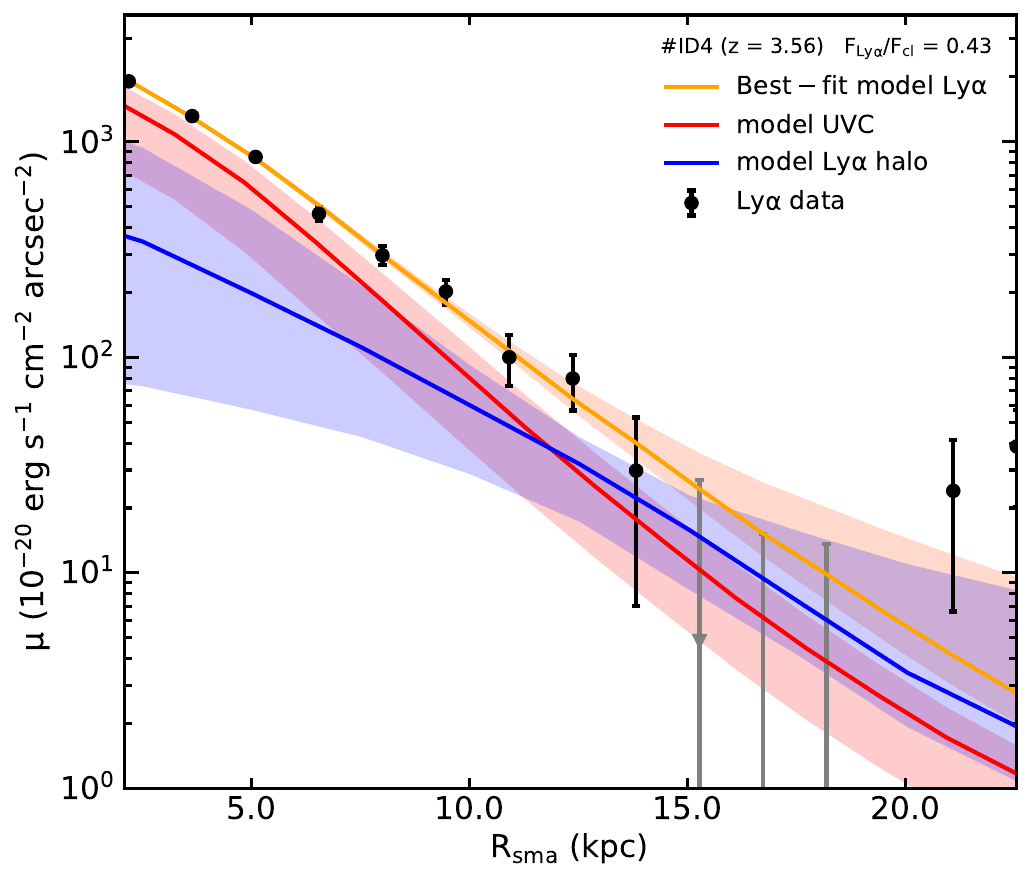}}
\hfill
\subfigure{\includegraphics[width = 2.3in, clip, trim = 0cm 0cm 0cm 0cm]{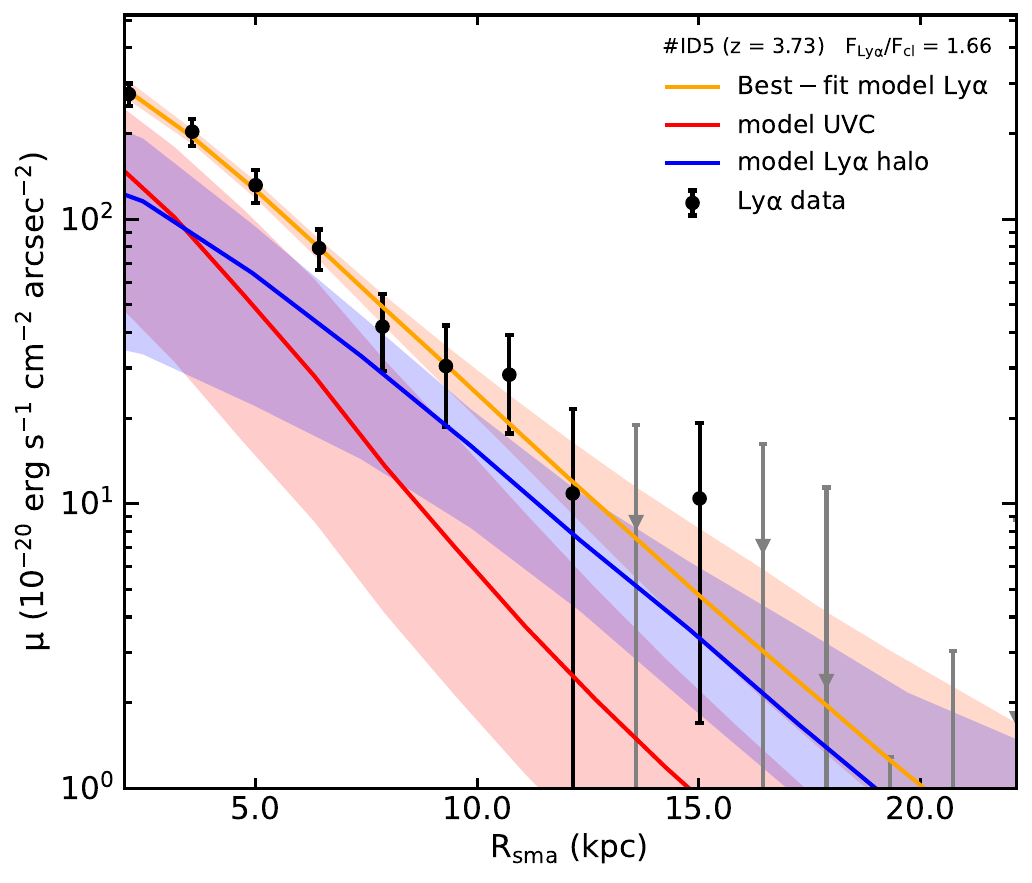}}
\hfill
\subfigure{\includegraphics[width = 2.3in]{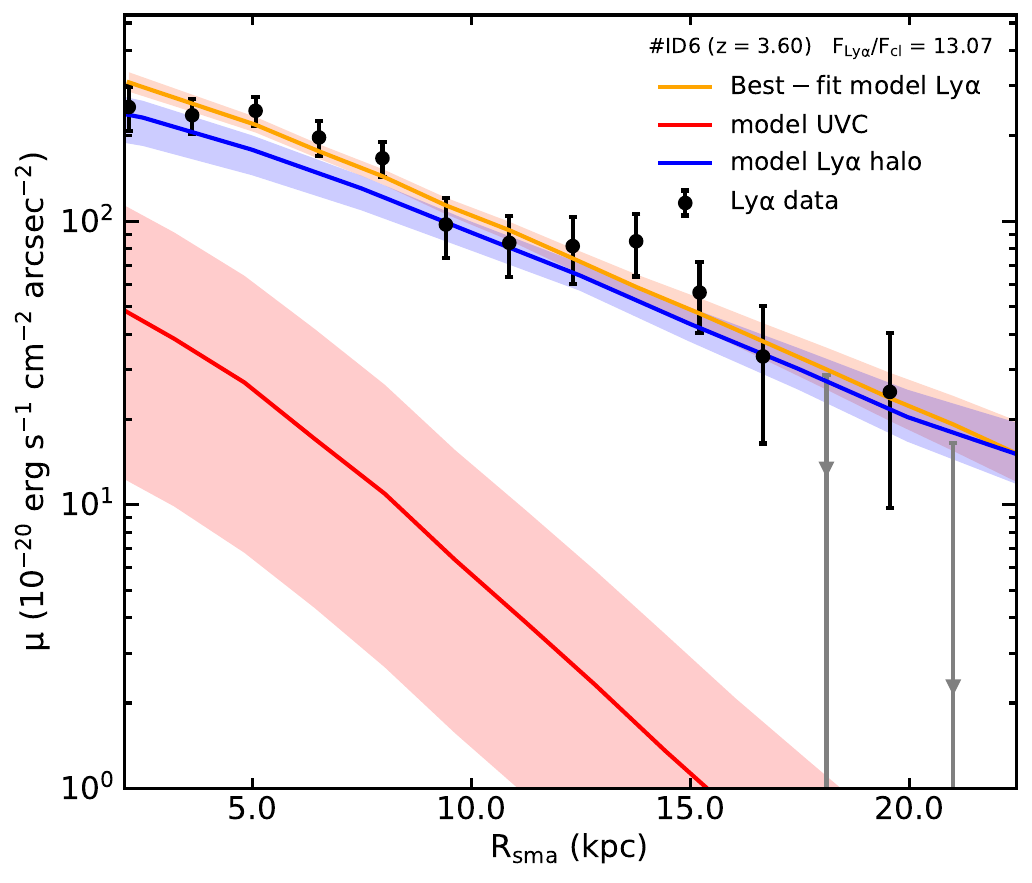}}\\

\subfigure{\includegraphics[width = 2.3in, clip, trim = 0cm 0cm 0cm 0cm]{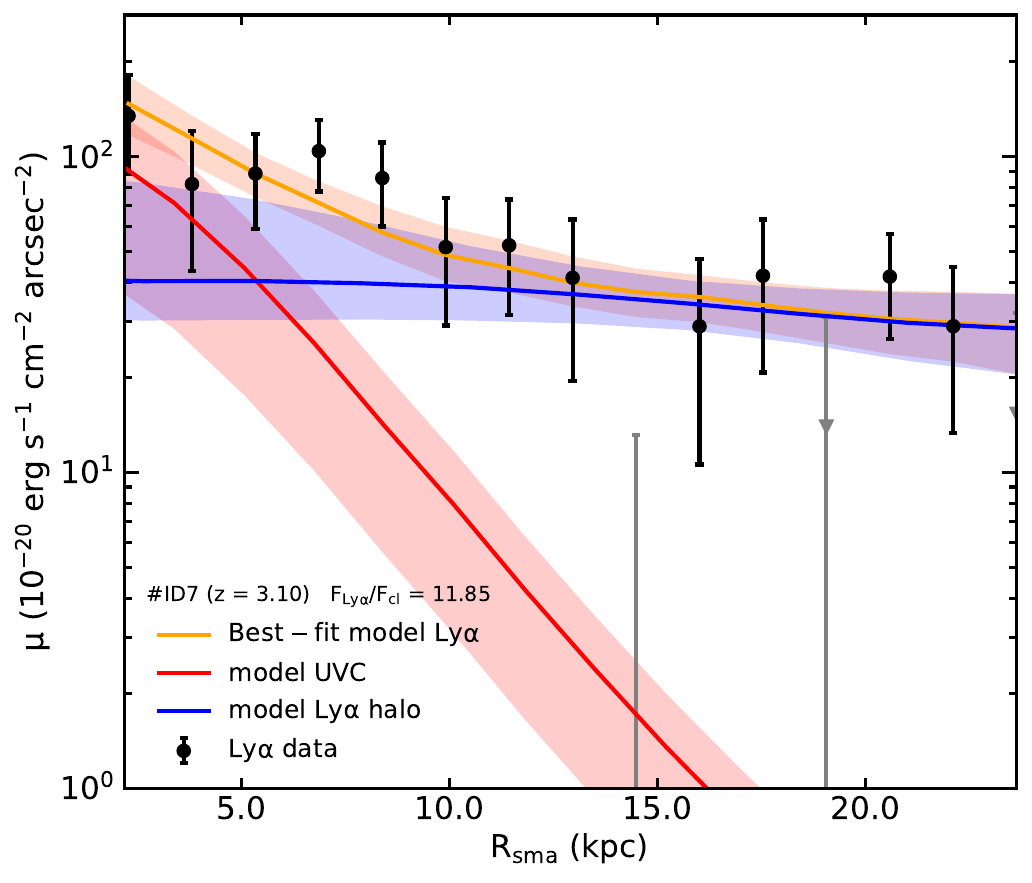}}
\hfill
\subfigure{\includegraphics[width = 2.3in, clip, trim = 0cm 0cm 0cm 0cm]{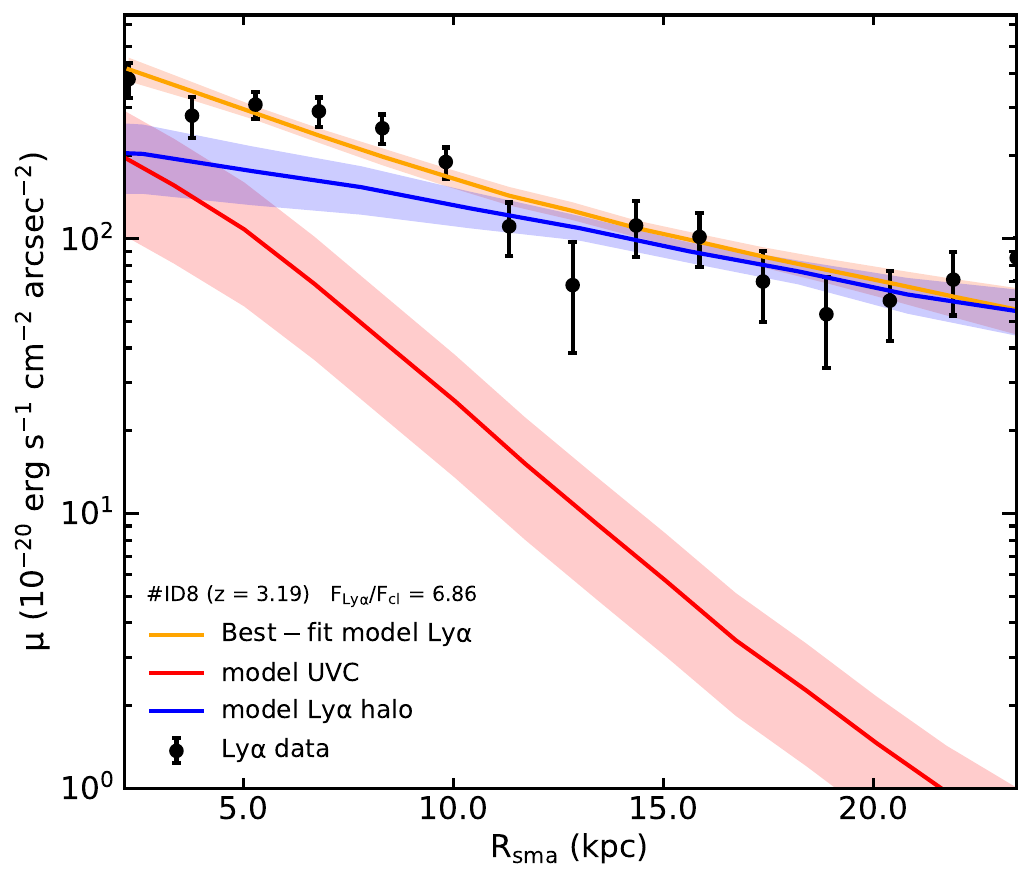}}
\hfill
\subfigure{\includegraphics[width = 2.3in]{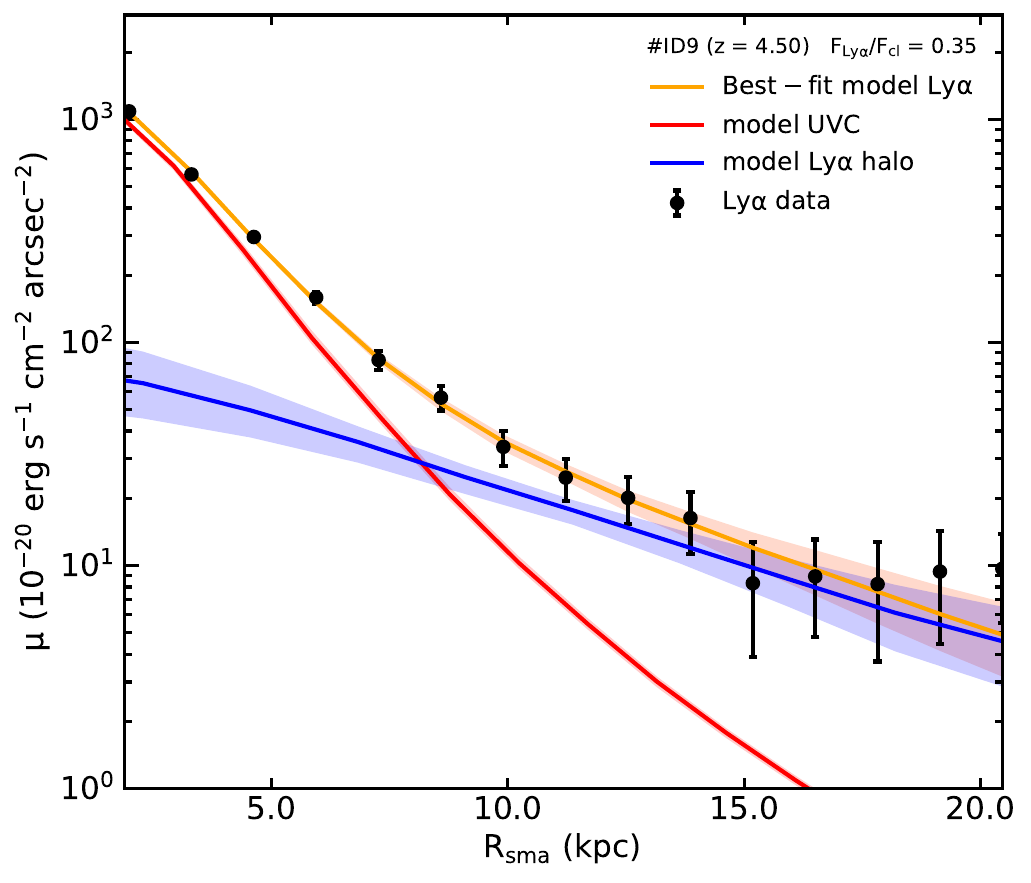}}\\

\subfigure{\includegraphics[width = 2.3in, clip, trim = 0cm 0cm 0cm 0cm]{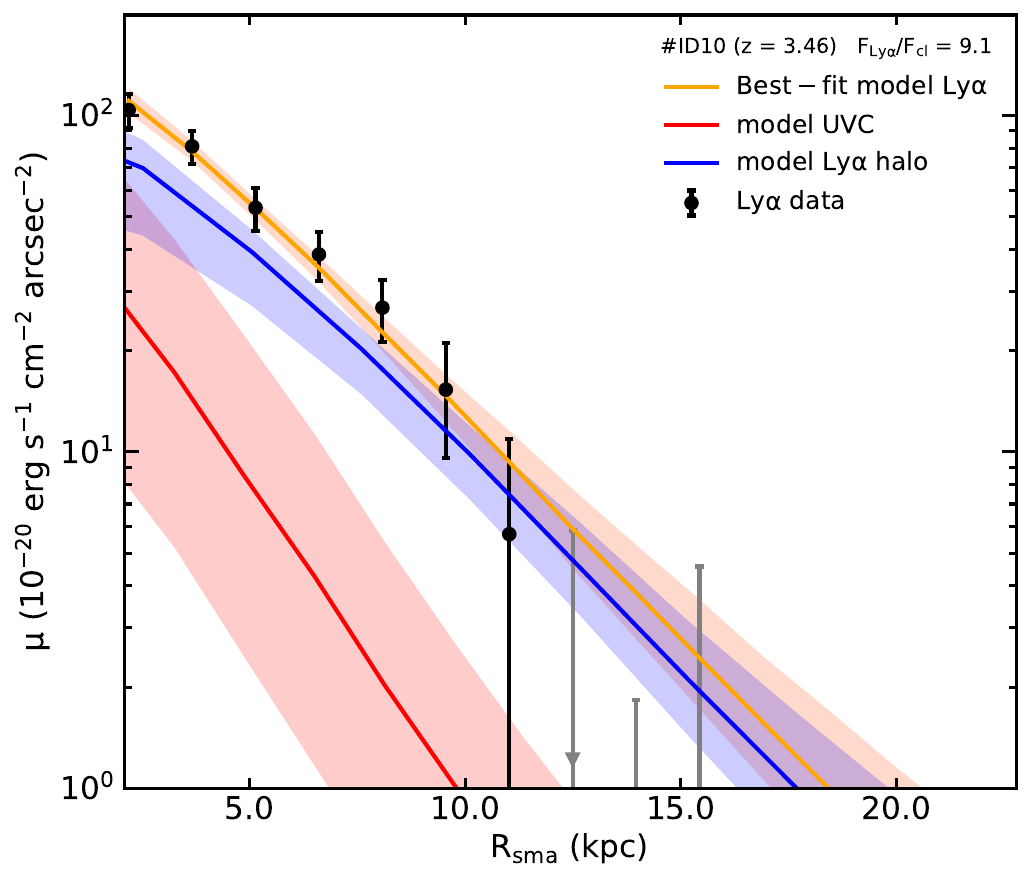}}
\hfill
\subfigure{\includegraphics[width = 2.3in, clip, trim = 0cm 0cm 0cm 0cm]{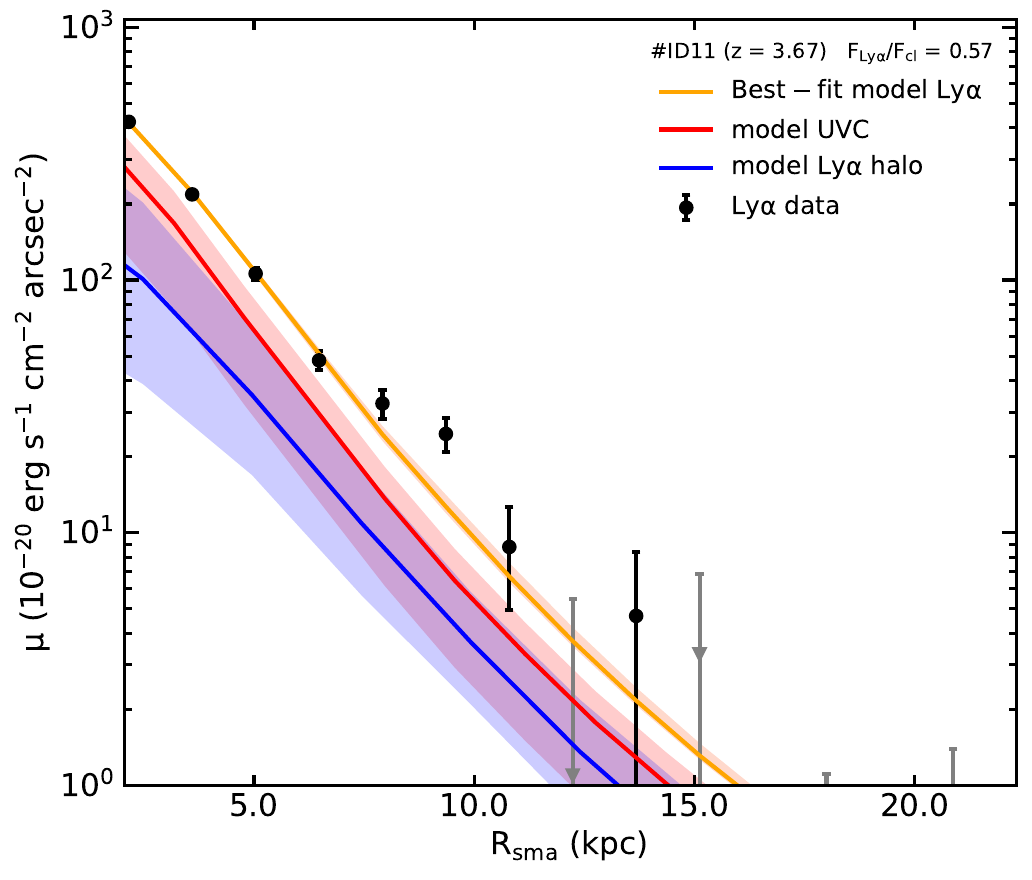}}
\hfill
\subfigure{\includegraphics[width = 2.3in]{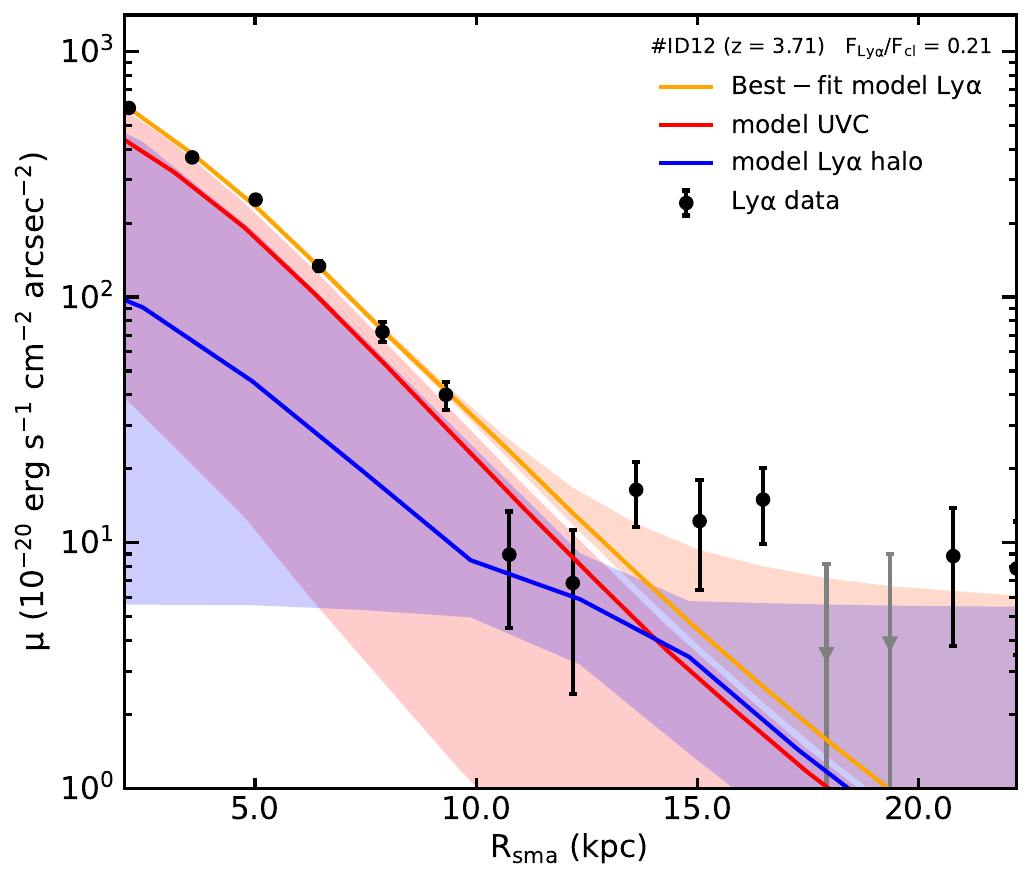}}\\
\caption{Two-component decomposition of the $\rm Ly\alpha$ surface brightness profiles (black points). The red curve corresponds to the compact $\rm Ly\alpha$ component, whose scale length is fixed to that measured from the HST UV continuum (UVC) image, while the shaded red region denotes the associated $1\sigma$ MCMC uncertainty. The blue curve represents the extended $\rm Ly\alpha$ halo component. The yellow curve shows the combined best-fitting model, obtained by summing the compact and halo components. The inferred flux ratio between the total $\rm Ly\alpha$ emission and the compact UVC-tracing component, $\rm F_{\rm Ly\alpha}/F_{\rm cl}$, is reported in each panel.}
\label{fig:lya_halo_fitting}
\end{figure*}

To quantify the spatial extent of the sample galaxies in both the UV continuum and $\rm Ly\alpha$ emission, we utilize the measured one-dimensional surface-brightness profiles. As described in Sec.~\ref{sec:sb_profile}, the centroid, position angle, and ellipticity of each source were determined and subsequently fixed for the extraction of the elliptical radial profiles. These structural parameters were then adopted as fixed inputs when fitting the UVC and $\rm Ly\alpha$ images using the \texttt{Sersic2D()} model implemented in the \texttt{Astropy} Python package \citep{Astropy_Collaboration2022}.

\begin{equation}
    I(x, y) = I_{e}\exp \left\{ -b_{n} \left[\left(\frac{r(x, y)}{r_{e}}\right)^{1/n} - 1\right] \right\}
\label{eqn:sersic2D}    
\end{equation}

The \texttt{Sersic2D} package uses Eqn.~\ref{eqn:sersic2D} \citep{Sersic1963, Sersic1968} to generate a two-dimensional Sersic model. In our analysis, we fix the Sersic index to n=1, thereby adopting an exponential surface-brightness profile. We further fix the source centroid, position angle, and ellipticity to the values measured in Sec.~\ref{sec:sb_profile}. Consequently, only the amplitude ($I_{e}$) and effective radius ($r_{e}$), defined as the radius enclosing 50\% of the total flux, are treated as free parameters during the fitting procedure.

We create the 2D models of the exponential profile and then convolve them with the averaged MUSE PSF at the wavelength range of $\rm Ly\alpha$ narrow band images. We then extract the 1D profile using the same elliptical annuli as for the surface brightness profile. To extract the best-fit profile, we use forward modeling and likelihood maximization for each galaxy. UV continuum profiles are fitted with a single exponential vial allowing the amplitude and effective radius to be free. Parameter estimation was performed within a Bayesian framework using the \texttt{emcee} ensemble sampler \citep{Foreman-Mackey2013, Foreman-Mackey2019} for Markov Chain Monte Carlo sampling. The likelihood function was defined from the $\chi^{2}$ difference between the observed and model surface-brightness profiles, weighted by the radial uncertainties. We use flat priors to restrict the parameters to physically meaningful values while fitting and to avoid points with negative surface brightness or noise exceeding the flux.

The posterior distributions were used to derive median parameter values and corresponding 16th–84th percentile confidence intervals. For each object, the adopted size measurement corresponds to the median posterior effective radius, while the quoted uncertainty reflects the marginalized credible interval.

We then used these UV continuum profiles and scaled them using a constant scaling factor ($f$) calculated via minimum $\chi^{2}$ statistics (see Eqn.~\ref{eqn:uv_scale_factor}) to match the measured $\rm Ly\alpha$ profiles, as the surface brightness units were not the same.

\begin{equation}
    f = \frac{\sum_{i = 1}^{N} \frac{L_{i}U_{i}}{\sigma_{i}^{2}}}{\sum_{i = 1}^{N} \left(\frac{U_{i}}{\sigma_{i}} \right)^{2}}
    \label{eqn:uv_scale_factor}
\end{equation}

where L is the $\rm Ly\alpha$ data points with error $\sigma$ on them, and U represents the median fitted UV profile points to the UV continuum data. 

We compare the scaled UV continuum profiles with the scaled PSF profiles, both normalized using Eqn.~\ref{eqn:uv_scale_factor}, as shown in Figure~\ref{fig:Lya_halo_spec_rgb}. Following the approach of \cite{Wisotzki_etal2016}, we test the null hypothesis that the UV continuum profile reproduces the $\rm Ly\alpha$ surface brightness distribution. For each galaxy, we compute the corresponding p-value to assess the validity of this assumption. We find that for sources \#ID 1, 2, 3, 4, 6, 7, 8, 9, 11, and 12, the null hypothesis is rejected (p-value $<$ 0.05), indicating the presence of extended $\rm Ly\alpha$ emission beyond the UV continuum. Additionally, several galaxies exhibit clear deviations between the inner $\rm Ly\alpha$ and UV profiles, suggesting structural differences even in the central regions.

To measure the effective radius of the $\rm Ly\alpha$ emission, we first fit it with a single-exponential profile, as was done for the UV profiles. We then added a second exponential component to construct a two-component model that accounts for the halo contribution. For fitting the halo, we first constrained the inner profile using the effective radius obtained from the UV continuum fitting and set the prior to the 16th--84th percentile range around the median value. The other parameters, namely the inner-component amplitude, halo effective radius, and halo amplitude, were kept free irrespective of the p-value. We fit the two-component model to all galaxies and found that the inner regions are generally dominated by a UV-like profile, except for \#ID 6, 10, and 12. In the case of \#ID 12, the UV-informed fit does not adequately reproduce the observed $\rm Ly\alpha$ profile.

We calculate the $\rm Ly\alpha$ halo to UVC flux ratio using the median posterior values obtained from the MCMC fitting. For each galaxy, we adopt the median of the fitted parameter distributions as the best-fit values. Using these parameters, we then create separate two-dimensional exponential models for the compact UVC-like component and the extended $\rm Ly\alpha$ halo component.

The total flux of each component is calculated by summing the flux over all pixels in the corresponding model image. We then compute the ratio between the total flux in the extended $\rm Ly\alpha$ halo model and that in the central UVC-like model. This ratio gives an estimate of the relative contribution of the diffuse halo emission compared to the compact inner component, and allows us to quantify how much of the observed $\rm Ly\alpha$ emission is distributed in the extended halo.

\section{Morphological and \texorpdfstring{$\rm Ly\alpha$\ }\  Characteristics of Individual Galaxies} \label{sec:individual_morphology}

\textbf{\#ID1:} This galaxy exhibits a clumpy morphology consisting of three distinct stellar clumps, one of which is significantly brighter than the others. The $\rm Ly\alpha$ emission appears more extended along the minor axis of the system, particularly when traced to the $\rm 1\sigma$ surface brightness contour.

\textbf{\#ID2:} The galaxy consists of a bright stellar clump accompanied by a faint diffuse tail. The $\rm Ly\alpha$ emission profile appears double peaked, with the bright emission spatially coincident with the bright clump.

\textbf{\#ID3:} As shown in Figure~\ref{fig:Lya_halo_spec_rgb}, this source exhibits four distinct stellar clumps, one of which displays a PSF-like morphology suggestive of an AGN component. The source is identified as a member of the recently discovered Little Red Dot (LRD) population \citep{Matthee_etal2024}. The $\rm Ly\alpha$ emission profile is double peaked and spatially extended along the orientation of the stellar clumps, while remaining centered on the AGN location. This object is also the highest-redshift and most $\rm Ly\alpha$ luminous galaxy in our sample.

\textbf{\#ID4:} This galaxy consists of two distinct stellar clumps associated with strong $\rm Ly\alpha$ emission. The emission line profile is singly peaked with an observed FWHM of $\sim 367~\rm km~s^{-1}$.

\textbf{\#ID5:} The system contains two distinct stellar clumps, while the $\rm Ly\alpha$ emission appears spatially offset from the continuum emission. The line profile is singly peaked and relatively faint compared to the rest of the sample.

\textbf{\#ID6:} The $\rm Ly\alpha$ emission in this galaxy is centrally concentrated near the tail component of the system and exhibits a prominent red peak in the spectral profile.

\textbf{\#ID7:} This source exhibits the weakest $\rm Ly\alpha$ emission in the sample. The stellar morphology is characterized by a broad diffuse tail and a faint clump, near which the detected $\rm Ly\alpha$ emission appears concentrated.

\textbf{\#ID8:} This galaxy exhibits the broadest $\rm Ly\alpha$ emission line in the sample and may show a double-peaked profile, although the low signal-to-noise ratio prevents a firm classification. The $\rm Ly\alpha$ emission appears partially detached from the stellar component and exhibits a plume-like morphology near the head of the galaxy. Together with the large line width, this may indicate the presence of neutral gas outflows and resonant scattering in the circumgalactic medium.

\textbf{\#ID9:} This object is the second highest-redshift galaxy in the sample. The brightest stellar clump is located near the head of the tadpole structure, showing a chain-like morphology. Unlike \#ID3, no point PSF-like component suggestive of AGN activity is observed.

\textbf{\#ID10:} The galaxy exhibits an extended tail together with a prominent head-like stellar component. A nearby galaxy with a similar color is also present and may be physically associated with the system. However, the $\rm Ly\alpha$ emission appears spatially coincident primarily with the tadpole galaxy, with no significant clumpy emission detected around the neighboring source.

\textbf{\#ID11:} The $\rm Ly\alpha$ emission is observed to align more closely with an external stellar clump rather than the central region of the host galaxy. Additionally, the spectral profile shows evidence for a faint blue peak component.

\textbf{\#ID12:} This source exhibits three distinct stellar clumps arranged in a chain-like morphology. The $\rm Ly\alpha$ emission is concentrated near the central clump and may exhibit a triple-peaked spectral structure.

\section{Summary \& Discussion} 
\label{sec:result_and_discussion}

We investigate the spatial distribution of $\rm Ly\alpha$ and UV continuum emission in a sample of tadpole/chain galaxies while explicitly accounting for their elongated stellar morphologies. After matching the HST UV continuum images to the MUSE point-spread function (PSF), we extracted surface brightness profiles using elliptical annuli aligned with the stellar morphology. The same procedure was applied to the $\rm Ly\alpha$ narrowband images to enable a consistent comparison between the stellar and $\rm Ly\alpha$ emission.

The model parameters were constrained using MCMC sampling, with only the amplitudes and scale lengths allowed to vary (see Figure~\ref{fig:lya_halo_fitting} for fitting). Although the stellar continuum exhibits highly elongated tadpole/chain-like morphologies, the $\rm Ly\alpha$ emission generally appears substantially more symmetric (see top left panel in Figure~\ref{fig:uv_lya_size_vs_prop}). This is evident from the $\rm Ly\alpha$ contours overlaid on the JWST/NIRCam RGB images (Fig.~\ref{fig:tpg_lya_spec}), where the diffuse emission often extends beyond the stellar distribution and exhibits a more circular morphology. 

Figure~\ref{fig:sample_comparison} compares the cumulative distributions of Ly$\alpha$ halo scale lengths and Ly$\alpha$ line widths for the tadpole galaxy sample with those of the much larger sample of Ly$\alpha$ emitters presented by \citet{Leclercq_etal2017}. The tadpole galaxies show a range of halo sizes and line widths that is broadly consistent with the general LAE population. While a few tadpole systems occupy the larger halo-size and broader-line-width regime, the overall distributions show substantial overlap with those of the comparison sample. Due to the limited size of our sample, we do not attempt to draw strong statistical conclusions regarding possible differences in spectra and sizes of todpole/chain-like galaxies compared to compact $\rm Ly\alpha$ emitting galaxies.

\begin{figure}
    \centering
    \includegraphics[width=1\linewidth]{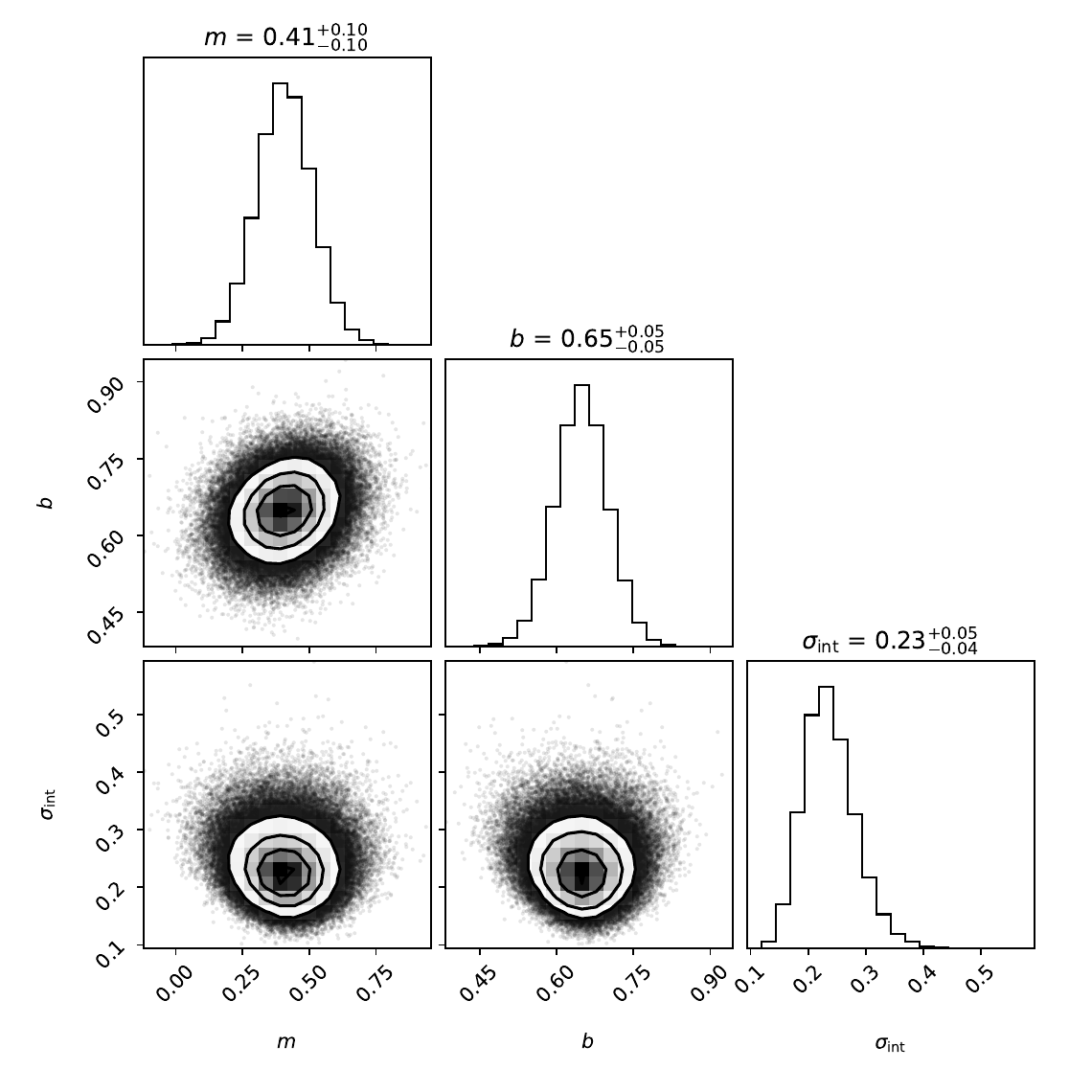}
    \caption{Posterior distribution of parameters used for fitting the datapoints in the top left plot of the Figure~\ref{fig:uv_lya_size_vs_prop} corresponding to Eqn~\ref{eqn:lya_uv_fit} with the scale lengths in log scale, \texttt{m} and \texttt{b} describes the slope and intercept respectively, while the \texttt{$\sigma_{int}$} is the intrinsic scatter in the fit to the data.}
    \label{fig:tpg_lya_uv_corner}
\end{figure}

Figure~\ref{fig:uv_lya_size_vs_prop} shows the relation between the $\rm Ly\alpha$ and UV continuum scale lengths. With their intrinsically elongated morphologies, the galaxies in our sample naturally extend the previously observed size relation toward larger UV scale lengths. Despite their highly elongated stellar structures, the galaxies continue to follow the trend in the $\rm Ly\alpha$ -- UV size plane, indicating that the $\rm Ly\alpha$ emission remains systematically more extended than the UV continuum and scales with UV size. To quantify this relationship, we fit the data in logarithmic scale lengths using an MCMC approach based on Eqn~\ref{eqn:lya_uv_fit}. The resulting posterior distributions of the fitted parameters are shown in the corner plot presented in Figure~\ref{fig:tpg_lya_uv_corner}.

\begin{equation}
    log(R_{eff}(Ly\alpha)) = (0.41^{+0.10}_{-0.10}) log(R_{eff}(UV)) + (0.65^{+0.05}_{-0.05})
    \label{eqn:lya_uv_fit}
\end{equation}

Extended $\rm Ly\alpha$ halos are detected around the majority of the galaxies in our sample. In most systems, the calculated halo fractions are less than unity (see Figure~\ref{fig:lya_halo_fitting}), indicating that a significant fraction of the observed $\rm Ly\alpha$ emission remains associated with the central stellar component. However, several galaxies display complex morphologies, including spatial offsets between the stellar and $\rm Ly\alpha$ emission. In particular, \#ID 8 exhibits detached $\rm Ly\alpha$ emission accompanied by a broad line profile, which may be indicative of large-scale gas motions or outflow-driven scattering.

We further examine the dependence of the $\rm Ly\alpha$ halo scale length on several other observed galaxy properties, including $\rm Ly\alpha$ luminosity, line FWHM, and rest-frame equivalent width. No strong correlations are identified, consistent with previous studies of high-redshift Ly$\alpha$ emitters. The measured halo scale lengths occupy a similar parameter space to those reported in \cite{Wisotzki_etal2016}, suggesting that tadpole galaxies do not form a distinct population in terms of their halo sizes alone.

Nevertheless, despite the similarities in halo scale lengths, the pronounced differences between the stellar and $\rm Ly\alpha$ morphologies, together with the high incidence of double-peaked $\rm Ly\alpha$ profiles, which can be the result of the selection criterion and the sample size, and frequent spatial offsets, suggest that resonant scattering plays an important role in redistributing Ly$\alpha$ photons in these systems. The disturbed, clumpy structures characteristic of tadpole galaxies may therefore influence the pathways by which Ly$\alpha$ photons escape and propagate into the circumgalactic medium.
 
\FloatBarrier
\bibliography{refrence}{}
\bibliographystyle{aasjournal}

\end{document}